\newcommand*{\ATLASLATEXPATH}{latex/}
\newcommand{\hd}{\hphantom{0}}

\newcommand{\hp}{\hphantom{.}}

\pdfoutput=1
\documentclass[cernpreprint, texlive=2016, UKenglish]{\ATLASLATEXPATH atlasdoc}

\usepackage{\ATLASLATEXPATH atlaspackage}
\usepackage{\ATLASLATEXPATH atlasbiblatex}
\usepackage{\ATLASLATEXPATH atlasphysics}

\graphicspath{{logos/}{figures/}}

\DeclareOldFontCommand{\bf}{\normalfont\bfseries}{\mathbf}
\DeclareOldFontCommand{\rm}{\normalfont\rmfamily}{\mathrm}

\usepackage{EXOT-2016-22-PAPER-defs}
\usepackage{algorithm2e}
\usepackage{multirow}

\AtlasTitle{Search for Higgs boson decays to beyond-the-Standard-Model light bosons in four-lepton events with the ATLAS detector at $\sqrt{s}=13$~\TeV}

\author{The ATLAS Collaboration}

\AtlasRefCode{EXOT-2016-22}

\PreprintIdNumber{CERN-EP-2017-293}

\AtlasJournal{JHEP}

\AtlasJournalRef{JHEP 06 (2018) 166} 

\AtlasDOI{10.1007/JHEP06(2018)166}

\AtlasAbstract{
  A search is conducted for a beyond-the-Standard-Model boson using events  
  where a Higgs boson with mass $125~\GeV$ decays to four leptons ($\ell =$ $e$ 
  or $\mu$). This decay is presumed to occur via an intermediate state which 
  contains one or two on-shell, promptly decaying bosons: $H\rightarrow 
  ZX/XX \rightarrow 4\ell$, where $X$ is a new vector 
  boson $Z_d$ or pseudoscalar $a$ with mass between 1 and 60~\GeV. The search 
  uses $pp$ collision data 
  collected with the ATLAS detector at the LHC with an integrated luminosity of 
  36.1~fb$^{-1}$ at a centre-of-mass energy $\sqrt{s}=13$~\TeV. No 
  significant excess of events above Standard Model background predictions is observed; therefore,  
upper limits at 95\% confidence level are set on model-independent fiducial 
cross-sections, and on the Higgs boson decay branching ratios to vector and 
pseudoscalar bosons in two benchmark models.

}

\hypersetup{pdftitle={ATLAS document},pdfauthor={The ATLAS Collaboration}}

\begin{document}

\maketitle

\tableofcontents

\section{Introduction}
\label{sec:intro}
Following the discovery of the Higgs boson by the ATLAS and CMS 
collaborations~\cite{ATLAS_Higgs,CMS_Higgs} at the Large Hadron Collider (LHC), 
a comprehensive programme of measurements of the properties of this particle 
is underway.
These measurements could uncover deviations from the expected branching ratios for the decays of a Standard Model (SM) Higgs boson or allow for the possibility of decays into non-SM particles.
Existing measurements constrain the non-SM or ``exotic'' branching ratio of the 
Higgs boson to less than approximately $30\%$ at $95\%$ confidence level 
(CL)~\cite{Aad:2015gba,Khachatryan:2014jba,Khachatryan:2016vau}. 

Exotic Higgs boson decays have been proposed as a way to search for evidence of new 
physics. Due to the extremely narrow decay width of the Higgs boson predicted 
by the SM, the addition of even a small coupling to 
a new light state could open up sizeable new decay modes. In addition, 
new particles may couple preferentially to the Higgs boson since it provides a 
possible ``portal'' for hidden-sector particles to interact with 
SM particles~\cite{Shrock:1982kd,Strassler:2006im,Schabinger:2005ei,Patt:2006fw}. 
Such decays are predicted by many theories of physics beyond the SM. 
For example, they are predicted in theories with a hidden (``dark'') 
sector~\cite{Fayet:2004bw,Finkbeiner:2007kk,ArkaniHamed:2008qn,Dudas:2012t1,Curtin:2014cca,Curtin:2013fra,Davoudiasl:2013aya,BNL2012dark,wells2008find,gopalakrishna2008higgs} 
and in those with an extended Higgs sector such as the Next-to-Minimal 
Supersymmetric Standard Model (NMSSM)~\cite{Dobrescu:2000yn,Ellwanger:2003jt,Dermisek:2005ar,Chang:2008cw,Morrissey:2008gm}. 
They are also predicted in several models of dark 
matter~\cite{Silveira:1985rk,Pospelov:2007mp,Draper:2010ew,Ipek:2014gua,Martin:2014sxa,Clowe}, 
models that explain astrophysical observations of positron 
excesses~\cite{Adriani:2008zr,ATIC:2008t1,Aguilar:110.141102}, models with a 
first-order electroweak phase transition~\cite{Profumo:2007wc,Blinov:2015sna}, 
and theories with neutral 
naturalness~\cite{Burdman:2006tz,Craig:2015pha,Curtin:2015fna}. 
The processes under study here are referred to as $pp \to H \to ZX/XX \to 4\ell$, with $Z$ being the SM $Z$ boson and with $X$ 
representing a possible new vector boson $Z_d$ or a new pseudoscalar boson $a$.
Section~\ref{sec:theory} provides an introduction to the theoretical background 
and specific models examined in this paper.

The search uses $pp$ collision data at a centre-of-mass energy $\sqrt{s} = 13$~TeV collected by the 
ATLAS detector (described in Section~\ref{sec:detector}) at the LHC in 
2015 and 2016 corresponding to an integrated luminosity of 36.1~fb$^{-1}$.
Same-flavour decays of the new particle to pairs of electrons and muons are 
considered, giving rise to the $4e$, $2e2\mu$, and $4\mu$ final states for particles in 
the mass range from $15~\GeV$ to $ m_H/2$, where $m_H=125$~\GeV.
For lower masses, targeting the range from $1~\GeV$ to $ 15~\GeV$, only the $4\mu$ 
final state is explored. Final states including $\tau$ leptons are not 
considered in either mass range.
The event reconstruction is discussed in Section~\ref{sec:evtsel}.

The search for $H \to ZX \to 4\ell$ in an $X$ mass range between 15~GeV and 
55~GeV is covered in Section~\ref{sec:ZZdAnalysis}, while the $H\rightarrow XX 
\rightarrow 4\ell$ searches are included in 
Sections~\ref{sec:ZdZdHighMassAnalysis}~and~\ref{sec:ZdZdLowMassAnalysis} for 
$15~\GeV < m_{X} < 60~\GeV$ and~$1~\GeV < m_{X} < 15~\GeV$, respectively.\footnote{The reason for the two ranges being
different is that in the $H\rightarrow ZX\rightarrow 4\ell$ search the mass
distributions of the $X$ and the $Z$~bosons begin to overlap significantly
for values larger than 55~\GeV, thus inhibiting unambiguous identification of
the $Z$ and the new bosons. This is not the case in the $H\rightarrow XX 
\rightarrow 4\ell$ search where a $Z$ veto is applied.}
Model interpretations and discussions are presented in 
Section~\ref{sec:Discussion_on_Interpretation}.
Finally, the conclusions of the search are presented in 
Section~\ref{sec:conclusion}.

This paper builds on the previous work of Ref.~\cite{EXOT-2013-15}, in which a 
similar analysis is reported with data collected at $\sqrt s = 8$~\TeV.

\section{Benchmark models}
\label{sec:theory}

Two well-motivated benchmark models that predict exotic decays to light 
beyond-the-Standard-Model (BSM) bosons are summarised below, 
and are used later in this paper when interpreting the results.
In the first BSM benchmark model, the SM is extended with a dark-sector $U(1)$ group, 
denoted $U(1)_d$, leading to the appearance of a BSM vector 
boson, $Z_d$. In the second BSM benchmark model, there are two Higgs doublets 
and an additional singlet scalar field (2HDM+S). This leads to the appearance 
of a BSM pseudoscalar boson, $a$.
The $Z_d$ boson and the $a$ pseudoscalar could each comprise the intermediate state in the decays 
$H \rightarrow ZX \rightarrow 4\ell$ and $H \rightarrow XX \rightarrow 
4\ell$, where the first benchmark model is considered for a higher mass range 
and the second for a lower mass range.

\subsection{Vector-boson model}
Hidden- or dark-sector states appear in many extensions to the 
SM~\cite{Fayet:2004bw,Finkbeiner:2007kk,ArkaniHamed:2008qn,Dudas:2012t1,Curtin:2014cca,Curtin:2013fra,Davoudiasl:2013aya,BNL2012dark,wells2008find,gopalakrishna2008higgs,Alexander:2016aln}.
 The dark-sector states allow a theoretically plausible route for generation of 
the particle content necessary to account for the astrophysical evidence 
of dark matter. For example, fermionic dark-matter candidates~\cite{Clowe} or dark-sector
couplings to normal matter might explain astrophysical observations of positron 
excesses~\cite{Adriani:2008zr,ATIC:2008t1,Aguilar:110.141102}.

A dark sector is introduced with an additional $U(1)_d$ dark gauge 
symmetry~\cite{Curtin:2014cca,Curtin:2013fra,Davoudiasl:2013aya,BNL2012dark,wells2008find,gopalakrishna2008higgs},
 coupled to the SM through kinetic mixing with the hypercharge gauge 
field~\cite{Galison:1983pa,Holdom:1985ag,Dienes:1996zr}. The gauge boson of the 
symmetry is the $\Zp$ vector boson. In this hypercharge portal scenario, the 
kinetic mixing parameter $\epsilon$ controls the coupling strength of the dark 
vector boson to SM particles, which in turn determines the lifetime of the 
$\Zp$ boson. The branching ratios of the $\Zp$ are independent of the kinetic 
mixing strength and are instead determined by the gauge coupling. This coupling 
leads to a significant fraction of decays ($\approx 15\%$) to pairs of 
electrons or muons. For $\Zp$ masses between 1~\GeV~and 60~\GeV, the decay would 
be prompt (relative to the vertex resolution of the ATLAS detector) for 
$\epsilon \gtrsim 10^{-5}$~\cite{Curtin:2014cca}. For smaller values of 
$\epsilon$, the displaced decays provide a unique signature, which has been 
previously searched for with the ATLAS detector in 
8~\TeV~collisions~\cite{EXOT-2013-22}. For $\Zp$ masses below a few~\GeV~and 
small values of $\epsilon$, the decay products would be highly collimated and 
require a special analysis~\cite{EXOT-2014-09}. Another possibility 
involves a mass mixing between the $Z$ boson and $\Zp$, facilitating the decay 
of the $\Zp$ to SM particles. In this mechanism, the strength of the mixing is 
determined by mass mixing parameter 
$\delta$~\cite{Davoudiasl:2013aya,BNL2012dark}.  %

If the $U(1)_d$ symmetry is broken by the introduction of a dark Higgs boson, there could also be mixing between the SM Higgs boson and the dark Higgs boson~\cite{Curtin:2014cca,Curtin:2013fra,Davoudiasl:2013aya,BNL2012dark,wells2008find,gopalakrishna2008higgs}.
In this scenario, the Higgs portal coupling $\kappa$ controls the strength of 
the Higgs coupling to dark vector bosons.
The observed Higgs boson would be the lighter one of an extended Higgs sector 
and could also decay into dark-sector particles.

 For the processes studied in this paper, the decay $H\rightarrow ZZ_d$ probes the 
 parameter space of $\epsilon$ and $m_{Z_d}$, and does not depend on the 
 presence of mixing between the SM Higgs boson and the dark-sector Higgs boson, 
 $\kappa$. However, this BSM signal is indistinguishable from SM $H\rightarrow 
 ZZ^*$ on an event-by-event basis, and therefore must emerge as a resonance 
 in the dilepton mass above this background process. 
 The SM background to the $H\rightarrow Z_dZ_d$ process, however, is more 
 easily separated from the signal. This feature makes the latter channel potentially 
 sensitive to much smaller values of kinetic mixing, where the only requirement 
 is that the kinetic mixing must be large enough for the $Z_d$ to decay promptly. 
 However, this process depends on the presence of mixing between the SM Higgs 
 boson and dark-sector Higgs boson, and therefore probes the parameter space of 
 $\kappa$ and $m_{Z_d}$.

\begin{figure}[!htbp]
\begin{center}
\includegraphics[width=0.48\textwidth]{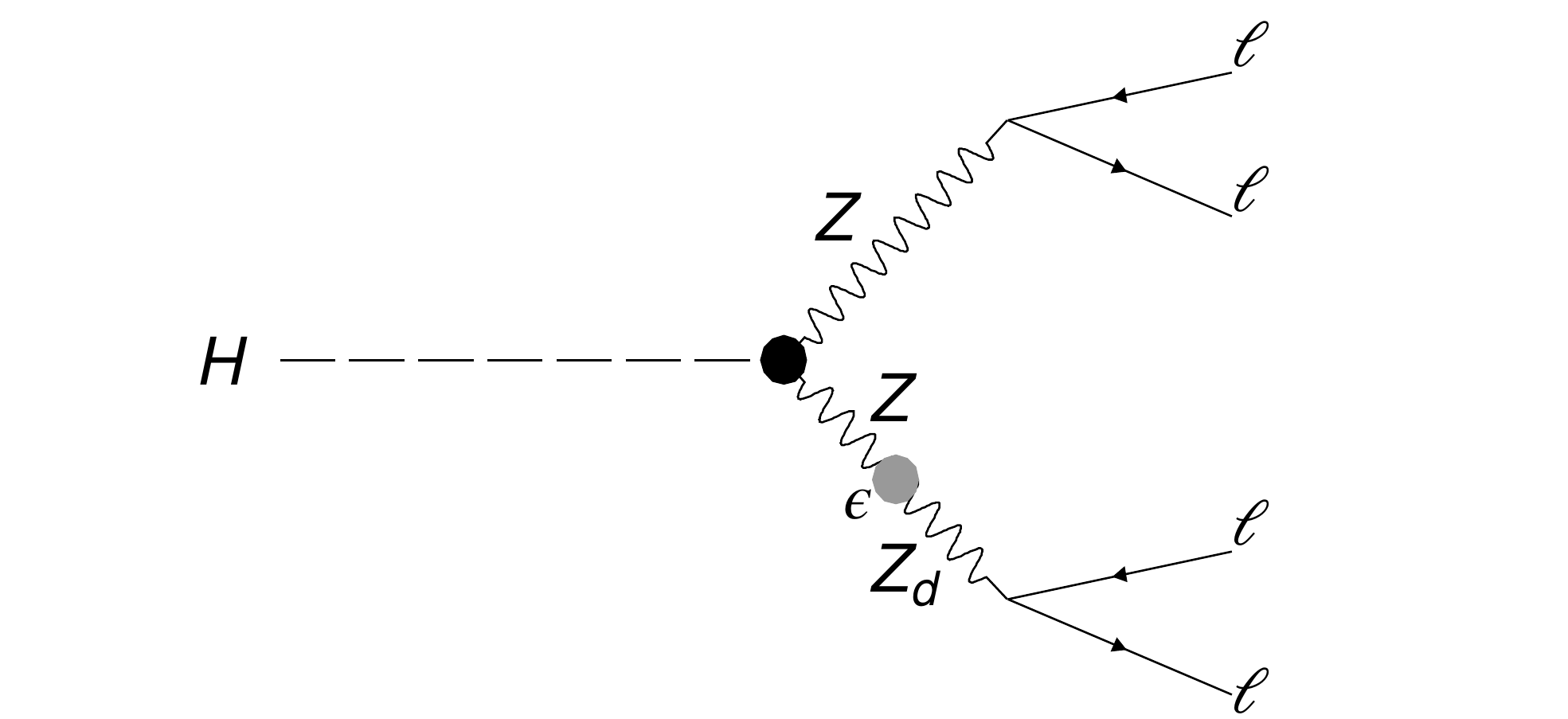}
\includegraphics[width=0.33\textwidth]{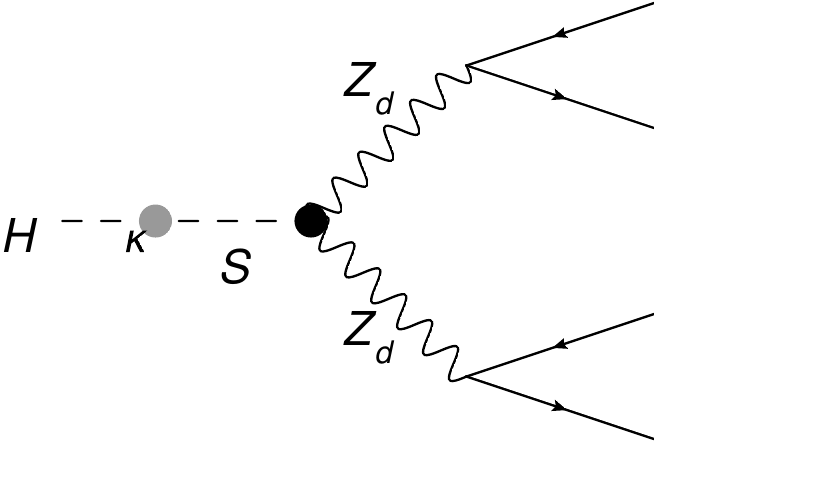}
\caption{Exotic Higgs boson decays to four leptons induced by intermediate dark vector bosons via (left) the hypercharge portal and (right) the Higgs portal, where $S$ is a dark Higgs boson~\cite{Curtin:2014cca}. The $\Zp$ gauge boson decays to SM particles through kinetic mixing with the hypercharge field or through mass mixing with the \Z~boson. The $HZZ_d$ vertex factor is proportional to $\epsilon$ whereas the $HZ_dZ_d$ vertex factor is proportional to $\kappa$.
\label{fig:ZpFeynman}}
\end{center}
\end{figure}

 Feynman diagrams of both processes are shown in Figure~\ref{fig:ZpFeynman}. These processes are 
 included in the Hidden Abelian Higgs Model (HAHM) that is used in this paper 
 as the benchmark vector-boson model~\cite{Curtin:2014cca}.

The presence of the dark sector could be inferred either from deviations from 
the SM-predicted rates of Drell--Yan (DY) events or from Higgs boson decays 
through exotic intermediate states.
Model-independent upper bounds from electroweak constraints on the kinetic 
mixing parameter, $\epsilon$, below 0.03 are reported in 
Refs.~\cite{Curtin:2014cca,Hook:2010tw,Pospelov:2008zw} for dark vector bosons with
masses between 1~\GeV~and 200~\GeV.
Upper bounds on the kinetic mixing parameter based on searches for dilepton 
resonances, $pp \to\Z_d \to \ell\ell$, below the $Z$~boson mass, are found to 
be in range of $0.005$--$0.020$ for dark vector bosons with masses between 20~\GeV~ 
and 80~\GeV~\cite{Hoenig:2014dsa}.
In the mass range of 10~\MeV--10~\GeV, values of $\epsilon$ above $\sim 10^{-3}$ are ruled out~\cite{Batley:2009aa,Lees:2012ra,PhysRevLett.59.755,PhysRevD.38.3375,PhysRevLett.67.2942,Adler:2004hp}.%
The experiments at the LHC are the only ones sensitive to the production of 
Higgs bosons, and this makes possible the search for the presence of a Higgs portal 
presented here. Constraints on the Higgs mixing parameter 
$\kappa$ are probed through the $H \rightarrow Z_dZ_d \rightarrow 4\ell$ search 
while constraints on the kinetic mixing parameter and the mass-mixing parameter $\delta$  
can be obtained through the $H \rightarrow \Z Z_d \rightarrow 4\ell$ search.

\subsection{Pseudoscalar-boson model}
Another possibility to extend the SM with a hidden sector is to consider two-Higgs-doublet models extended by one complex scalar singlet field 
(2HDM+S)~\cite{Curtin:2013fra}.

Two-Higgs-doublet models predict two charged scalars ($H^{\pm}$), two neutral scalars ($H$, $\mathbb{H}$) and one neutral pseudoscalar ($A$).
The real mass eigenstate $H$ is considered to be the observed Higgs boson, 
while other states are taken to be heavy in the decoupling limit to ensure that 
highly non-standard Higgs decays (e.g.\ involving CP-violation) which are 
significantly constrained by existing data, are 
avoided~\cite{Gunion:1997aq,Belanger:2013xza}.
The scalar singlet added to 2HDM only couples to the two Higgs complex fields in the potential and has no Yukawa couplings.
Therefore, all of its couplings to SM fermions are acquired through mixing of 
the scalar field with the Higgs complex fields, which needs to be small to 
preserve the SM nature of the Higgs sector.

With these assumptions, the decay $H\rightarrow aa$ is allowed, where $a$ is a light pseudoscalar mass eigenstate mostly composed of the imaginary part of the singlet field.\footnote{The pseudoscalar state is $a = \cos\theta_a S_I + \sin\theta_a A$, where $\theta_a \ll 1$ is a small mixing angle and $S_I$ is the imaginary part of the complex singlet field.}
The aforementioned constraints on two-Higgs-doublet models can be incorporated in the 2HDM+S by choosing a region of the 2HDM phase space not yet excluded, and giving the real and imaginary components of the singlet separate masses and small mixings to the Higgs doublets.
The branching ratios of $a$ into fermions are determined by the Yukawa 
couplings of $a$ to fermions, and lead to a rich decay 
phenomenology~\cite{Curtin:2013fra}, albeit with typically negligible branching 
ratio to pairs of electrons, and smaller branching ratios to pairs of muons 
than the dark vector bosons described in the previous section.
Among all the models predicting different decay possibilities, type~II
 are theoretically well motivated,\footnote{The right-handed states $d_R$ and $e_R$ couple to $H_{1}$, $u_R$ to $H_{2}$, where $H_{1}$ and $H_{2}$ are the two Higgs doublets. See Ref.~\cite{Curtin:2013fra} for more information.} since light pseudoscalars can correspond to the $R$-symmetry limit of the NMSSM~\cite{Dermisek:2006wr,Ellis:1988er}, which elegantly solves the $\mathrm{\mu}$-problem of the MSSM ~\cite{Giudice:1988yz} and greatly reduces the fine-tuning and little-hierarchy problems.
Furthermore, in the NMSSM the branching ratio for $H\rightarrow aa$ can be significant.
Type-II models can also predict a significant branching ratio for 
$a\rightarrow\mu\mu$, especially in the range $2m_{\mu}\ <\ m_{a}\ < 
2m_{\tau}$, with values ranging from~$10^{-2}$ to~$10^{-1}$ for some regions of 
the parameter space~\cite{Curtin:2013fra}.

Several searches for a Higgs boson decaying to electrons, muons, $\tau$ leptons or $b$-jets via two pseudoscalars have been performed at both the LHC and the Tevatron.
The D\O{} and ATLAS collaborations have searched for a signal of $H\to aa \to 2\mu 2\tau$ in the $a$~boson mass ranges $3.7 \leq m_a \leq 19~\GeV$~and~$3.7 \leq m_a \leq 50~\GeV$, respectively~\cite{Abazov:2009yi,Aad:2015oqa}.
The D\O{} and CMS collaborations have searched for the signature $H \to aa \to 
4\mu$ in the range $2 m_{\mu} \leq m_a \leq 2 
m_{\tau}$~\cite{Abazov:2009yi,Khachatryan:2015wka}.
The CMS collaboration has additionally searched for $H \to aa \to 4\tau, 2\mu 
2\tau, 2\mu 2b$ in the range $5~\GeV \leq m_a \leq 
62.5~\GeV$~\cite{Khachatryan:2017mnf} and the ATLAS collaboration for $H \to aa 
\to 4b$ in the range $20~\GeV \leq m_a \leq 60~\GeV$~\cite{Aaboud:2016oyb}.
These searches have led to limits on the branching ratio of the Higgs boson 
decaying to $aa$, scaled by the ratio of the production cross-section of the 
Higgs boson that is searched for to that predicted by the SM, 
$\sigma(H)/\sigma_{\text{SM}} \times \BR(H\rightarrow aa)$, 
between 1\% and 3\% for pseudoscalar-boson masses between 1~\GeV~and 3~\GeV~and 
between 10\% and 100\% for masses larger than 5~\GeV, assuming a 2HDM+S Type-II 
model with $\tan \beta = 5.0$.

\section{ATLAS detector}
\label{sec:detector}

The ATLAS experiment~\cite{PERF-2007-01} is a multi-purpose particle
physics detector with forward-backward symmetric cylindrical geometry
and a near 4$\pi$ coverage in solid angle.\footnote{ATLAS uses a
  right-handed coordinate system with its origin at the nominal
  interaction point (IP) in the centre of the detector and the
  $z$-axis along the beam pipe. The $x$-axis points from the IP to the
  centre of the LHC ring, and the $y$-axis points upward. Cylindrical
  coordinates $(r,\phi)$ are used in the transverse plane, $\phi$
  being the azimuthal angle around the $z$-axis. The pseudorapidity is
  defined in terms of the polar angle $\theta$ as
  $\eta=-\ln\tan(\theta/2)$. The transverse momentum \pt~and other transverse variables, 
are defined as the variables' component in the $x-y$ plane, 
the transverse energy \et is defined as $\sqrt{m^2+ \pt^2}$, where $m$ represents the mass of a considered object. 
The distance in the pseudorapidity--azimuthal-angle space is defined as dR 
or $\Delta R = \sqrt{(\Delta \eta)^2 + (\Delta \phi)^2}$.} The interaction point is surrounded by
an inner detector (ID) tracking system, a calorimeter system, and a muon spectrometer
(MS). The ID covers $|\eta| < 2.5$ and consists of a silicon pixel
detector, a silicon microstrip detector, and a transition radiation
tracker. The ID is surrounded by a thin superconducting solenoid
providing a 2 T axial magnetic field. One significant upgrade for
Run 2 is the presence of the insertable B-layer
(IBL)~\cite{ATLAS-TDR-19}, an additional pixel layer close to the
interaction point, which provides high-resolution measurements at small radius
to improve the tracking performance. The calorimeter system features a
high-granularity lead/liquid-argon (LAr) sampling calorimeter that
measures the energy and the position of electromagnetic showers within
$|\eta| < 4.9$. LAr sampling calorimeters are also used to measure
hadronic showers in the endcap ($1.5 < |\eta| < 3.2$) and forward
($3.1 < |\eta| < 4.9$) regions, while an steel/scintillator tile
calorimeter measures hadronic showers in the central region ($|\eta| <
1.7$). The MS surrounds the calorimeters and consists of three large 
superconducting air-core toroid magnets, each with eight coils, a
system of precision tracking chambers ($|\eta| < 2.7$), and fast
trigger chambers ($|\eta| < 2.4$). For Run 2 the ATLAS detector has a 
two-level trigger system. The first-level trigger (Level-1 trigger) is
implemented in hardware and uses a subset of the detector information
to reduce the accepted rate to 100 kHz. This is followed by a
software-based trigger (called high-level trigger) that reduces
the rate of events recorded to 1 kHz.

\section{Event reconstruction}
\label{sec:evtsel}
The three analyses presented in this paper all follow a similar event 
reconstruction and selection procedure. This section describes the basic event 
selection and lepton reconstruction requirements that are common to all three 
analyses. Table~\ref{tab:eventselection} summarises the event selection used in 
the three analyses that are described in further detail in 
Sections~\ref{sec:ZZdAnalysis}--\ref{sec:ZdZdLowMassAnalysis}.

Events are preselected in accord with trigger requirements and basic event 
requirements such as the existence of a reconstructed primary vertex~\cite{ATL-PHYS-PUB-2015-026}, which has the
largest sum of $\pt^{2}$ of the associated tracks. For each 
event, a selection is applied to the reconstructed final-state leptons. 
The event is required to have at least four leptons.
These leptons are combined into dileptons, and the dileptons are paired into 
quadruplets. Quadruplets are then filtered by selection criteria specific to 
each analysis, and a single quadruplet (with a specific dilepton pairing) is 
selected according to a ranking metric that favours pairings compatible with either 
a $ZX$ or $XX$ intermediate state, depending on the analysis. If there are no 
quadruplets in the event that meet the selection criteria then the event is 
discarded. Final event selections are based on properties of this selected 
quadruplet and the corresponding dilepton pair.

\subsection{Trigger and event preselection}

Events are preselected by single-lepton, dilepton, or trilepton 
triggers~\cite{TRIG-2016-01}, with a combined efficiency very close to 100\% 
(relative to the signal region events surviving all other event 
selections). Trigger thresholds were increased slightly throughout the run to 
compensate for increasing peak instantaneous luminosity delivered by the LHC. 
The lowest $\pt$ thresholds for the single-lepton triggers ranged from 
$24~\GeV$ to $26~\GeV$. Dielectron (dimuon) trigger 
thresholds ranged from $2\times12~\GeV$ ($2\times10~\GeV$) to 
$2\times17~\GeV$ ($22,8~\GeV$). Trielectron (trimuon) triggers had thresholds 
of $17,9,9~\GeV$ ($3\times 6~\GeV$). In the low-mass selection, only the 
muon-based triggers are used. The events must have at least one primary 
vertex~\cite{ATL-PHYS-PUB-2015-026} with two or more associated tracks with 
$\pt > 400$~\MeV~and satisfy cleaning criteria~\cite{ATLAS-CONF-2015-029} designed 
to reject events with excessive noise in the calorimeters.

\subsection{Lepton reconstruction}

\label{sec:object_selection}

An electron is reconstructed from a cluster of energy deposits in the 
electromagnetic calorimeter matched to a high-quality track in the ID. Its 
momentum is computed from the cluster energy and the direction of the track. 
Electrons are required to have $|\eta| < 2.47$ and $\pt > 7$~\GeV.
Electrons can be distinguished from other particles using several 
identification criteria that rely on the shapes of electromagnetic showers as 
well as tracking and track-to-cluster matching quantities. Following the 
description in Ref.~\cite{PERF-2016-01}, the output of a likelihood function 
taking these quantities as input is used to identify electrons, choosing the 
loose working point, but with the additional requirement of a hit presence in 
the innermost layer of the ID.\footnote{When no measurement is 
expected in the innermost layer of the pixel detector, the
requirement is transferred to the next-to-innermost pixel layer.
}

A muon is reconstructed by matching a track or track segment reconstructed in 
the MS to a track reconstructed in the ID~\cite{PERF-2015-10}. Its momentum is 
calculated by combining the information from the two systems and correcting for 
energy deposited in the calorimeters.
In regions of limited coverage by the MS ($|\eta| < 0.1$), muons can be 
reconstructed by matching ID tracks to calorimeter signals consistent with a 
minimum-ionising particle (calorimeter-tagged muons).
In regions outside the ID acceptance ($2.5 < |\eta| < 2.7$), muon 
reconstruction can also be extended by using tracks in the MS (stand-alone 
muons).
Reconstructed muons are required to pass the requirements of the loose working 
point to maximise the reconstruction efficiency while providing good-quality 
muon tracks~\cite{PERF-2015-10}.
Muons are required to have $|\eta| < 2.7$ and $\pt > 5$~\GeV. 
Calorimeter-tagged muons must have $\pt > 15$~\GeV.

Leptons are required to originate from the hard-scattering vertex, defined as the primary vertex in the pre-selection. 
The longitudinal impact 
parameter of each lepton track, calculated relative to the hard-scattering 
vertex and multiplied by $\sin\theta$ of the track, is required to be smaller 
than 0.5~mm. Furthermore, muons must have a transverse impact parameter 
calculated relative to the beam line smaller than 1~mm in order to reject 
muons originating from cosmic rays.
The significance of the transverse impact parameter calculated relative to 
the beam line is required to be less than three (five) for muons (electrons).
Stand-alone muons are exempt from all three requirements, as they do not have an 
ID track.

The leptons are required to 
be isolated from other particles using ID track information and calorimeter 
information.
The sum of the transverse energy $\Sigma \et$ of other topological 
clusters~\cite{PERF-2014-07} in the cone of $\Delta R=0.2$ around the electron (muon) is 
required to be less than $20\%$ ($30\%$) of the $\pt$ of the electron (muon).
The $\Sigma \pT$ of tracks within a variable-width cone of $\Delta 
R=\text{min}(0.2,10~\GeV/\pt)$ ($\Delta R<\text{min}(0.3,10~\GeV/\pT)$) of the 
electron (muon) must be less than $15\%$ of the $\pt$ of the electron (muon).
Contributions to the isolation cones from other leptons in the quadruplet are 
subtracted before applying the requirements.

Overlap removal is applied to avoid identifying the same detector signature as 
multiple electrons, muons or jets.
Electrons sharing an ID track with a selected muon are ignored, 
except if the muon is only calorimeter-tagged, in which case the muon is 
ignored instead. Electrons sharing their track or cluster in the 
calorimeter 
with a selected higher-\pt~electron are ignored.

\subsection{Definition of invariant-mass kinematic variables}
For all three analyses, the convention is adopted that $m_{12}$ and $m_{34}$ are 
the invariant masses of the two dileptons that make up a quadruplet, with the 
defining constraint that $|m_{12}-m_{Z}|<|m_{34}-m_{Z}|$, where $m_{Z}$ is the 
mass of the $Z$ boson\footnote{Put another way, $m_{12}$ is the
invariant mass of the dilepton that is closer to the Z boson mass, and $m_{34}$ 
is the invariant mass of the other dilepton in the quadruplet.}~\cite{1674-1137-38-9-090001}. Thus $m_{12}$ 
identifies the primary pair and $m_{34}$ is the secondary pair.

In the case of quadruplets formed from four electrons or four muons, alternate 
pairings of same-flavour opposite-sign leptons can be formed. The invariant 
masses of these alternate pairings are denoted by $m_{14}$ and $m_{32}$, where 
the 
positively charged lepton from the primary pair is paired with the negatively 
charged lepton from the secondary pair to compute $m_{14}$, and the positively 
charged lepton from the secondary pair is paired with the negatively charged 
lepton from the primary pair to compute $m_{32}$.

\subsection{Summary of analysis event selections}
Table~\ref{tab:eventselection} summarises the event selection used in the three 
analyses that are described in further detail in 
Sections~\ref{sec:ZZdAnalysis}--\ref{sec:ZdZdLowMassAnalysis}, and signal 
efficiencies of these selections with respect to a minimal fiducial volume are 
shown in figures~\ref{fig:efficiency_ZZd} and~\ref{fig:efficiency_ZdZd} of 
section~\ref{sec:fiducialLimits}.

\begin{table}[!htbp]
\caption{Summary of the event selection of the different analyses described
        in this paper. The quarkonia resonance masses $m_{J/\Psi}, m_{\Psi(2S)}, 
        m_{\Upsilon(1S)}$, and $m_{\Upsilon(3S)}$ are taken from
        Ref.~\cite{1674-1137-38-9-090001}.}

	\footnotesize
	\begin{center}
		
				\begin{tabular}{>{\arraybackslash}p{0.09\linewidth}|>{\centering\arraybackslash}p{0.27\linewidth}|>{\centering\arraybackslash}p{0.225\linewidth}|>{\centering\arraybackslash}p{0.255\linewidth}}
					
					\toprule
					& \textbf{$H\rightarrow ZX \rightarrow 4\ell$ \newline ($15~\GeV < m_{X} < 55~\GeV$)} &
                    \textbf{$H\rightarrow XX \rightarrow 4\ell$ \newline ($15~\GeV < m_{X} < 60~\GeV$)} &
                    \textbf{$H\rightarrow XX \rightarrow 4\mu$ \newline ($1~\GeV  < m_{X} < 15~\GeV$)} \\
					\hline
					\multirow{2}{1.0\linewidth}{\textsc{Quadruplet selection}} & \multicolumn{3}{l}{- Require at least one quadruplet of leptons consisting of two pairs of same-flavour opposite-sign leptons} \\
					& \multicolumn{3}{l}{- Three leading-\pt{} leptons satisfying $\pt > 20$~\GeV{}, 15~\GeV{}, 10~\GeV{}}\\
					& \multicolumn{3}{l}{- At least three muons are required to be reconstructed by combining ID and MS tracks in the $4\mu$ channel} \\
					\cline{2-4}
					& - Select best quadruplet (per channel) to be the one with 
					the (sub)leading dilepton mass (second) closest to the $Z$ 
					mass & 
					\multicolumn{2}{>{\centering\arraybackslash}p{0.55\linewidth}}{
					 Leptons in the quadruplet are responsible for firing at 
					 least 
					one trigger. In the case of multi-lepton triggers, all 
					leptons of 
					the trigger must match to leptons in the quadruplet} 
					\\
					& - $50~\GeV < m_{12} < 106~\GeV$ & \multicolumn{2}{l}{} \\
					& - $12~\GeV < m_{34} < 115~\GeV$ & \multicolumn{2}{l}{} \\
					& - $m_{12,34,14,32}>5 $~\GeV & \multicolumn{2}{l}{} \\
					\cline{2-4}
					& \multicolumn{2}{>{\centering\arraybackslash}p{0.55\linewidth}|}{$\Delta R(\ell,\ell')>0.10\ (0.20)$ for same-flavour (different-flavour) leptons in the quadruplet} & - \\
					\hline
					\textsc{Quadruplet ranking} &  Select first surviving quadruplet from channels, in the order:  $4\mu$, $2e2\mu$, $2\mu 2e$, $4e$ & \multicolumn{2}{c}{ Select quadruplet with smallest $\Delta m_{\ell\ell}=|m_{12}-m_{34}|$}  \\
					\hline
					\multirow{2}{1.0\linewidth}{\textsc{Event selection}} & \multicolumn{2}{c|}{$115~\GeV < m_{4\ell} < 130~\GeV{}$} & $120~\GeV < m_{4\ell} < 130~\GeV{}$ \\  
					\cline{2-4}
					& &\multicolumn{2}{c}{$m_{34}/m_{12} > 0.85$}\\
					& &\multicolumn{2}{c}{Reject event if:}\\
					& &\multicolumn{2}{c}{ $\mbox{ }$ $\mbox{ }$ $(m_{J/\Psi}-0.25~\GeV) < m_{12,34,14,32} < (m_{\Psi(2S)}+0.30~\GeV)$, or }\\
					& &\multicolumn{2}{c}{ $\mbox{ }$ $\mbox{ }$ 
					$(m_{\Upsilon(1S)}-0.70~\GeV) < m_{12,34,14,32} < 
					(m_{\Upsilon(3S)}+0.75~\GeV)$}\\
					\cline{3-4}
					& & $10~\GeV < m_{12,34} < 64~\GeV$ & $0.88~\GeV < m_{12,34} < 20$~\GeV{}\\
					& & $4e$ and $4\mu$ channels: $5~\GeV <m_{14,32} <75~\GeV$ & No restriction on alternative pairing\\
					\bottomrule
				\end{tabular}
	\end{center}
	\label{tab:eventselection}
\end{table}

\section{$H\rightarrow ZX \rightarrow 4\ell$ analysis}
\label{sec:ZZdAnalysis}
\subsection{Monte Carlo simulation}

\label{sec:ZZdAnalysis_MCsimulation}

Samples of events with \HZZpllll, where the Higgs boson with mass $m_H=125~\GeV$ was produced in the gluon--gluon fusion mode ($gg$F), were generated using the Hidden Abelian Higgs Model (HAHM)~\cite{wells2008find,gopalakrishna2008higgs}. The event generator \progname{MadGraph5\_aMC@NLO} v2.2.3~\cite{Alwall:2014hca} with the NNPDF23~\cite{Ball:2012cx}  parton distribution functions (PDFs) at leading order (LO) was used. \progname{Pythia8}~\cite{pythia2008} (v8.170)  with the A14 parameter set~\cite{cite:autwo} was used for the modelling of the parton shower, hadronisation and underlying event. Nine samples were generated in the range $15 \leq m_{Z_{d}} \leq 55$~\GeV~ with a 5~\GeV~step corresponding to different $Z_d$ mass hypotheses. The model parameters  $\epsilon$ and $\kappa$ were adjusted so that only \HZZpllll~decays were generated ($\epsilon \gg \kappa$). The samples were normalised using the SM Higgs boson production cross-section $\sigma_{\rm SM}(gg\rm{F}) = 48.58$~pb~and the $\BR(\HZZllll)=1.25\times 10^{-4}$ taken from Ref.~\cite{Heinemeyer:2013tqa}, as this branching ratio corresponds approximately to the upper limit set in the previous search~\cite{EXOT-2013-15}. Final states with $\tau$ leptons are not considered in this analysis and thus were not generated. %

The background processes considered for this search follow those used in the \HZZllll~measurement~\cite{HIGG-2016-25}, and consist of:

\begin{itemize}
\item[] \textbf{\HZZllll:}
The  Higgs boson production through $gg$F~\cite{Hamilton:2013fea}, vector-boson fusion (VBF)~\cite{Nason:2009ai}, and in association with a vector boson ($VH$)~\cite{Luisoni:2013kna}, was simulated using the \progname{Powheg-Box} v2 MC event generator~\cite{Nason:2004rx,Frixione:2007vw,Alioli:2010xd} with the PDF4LHC NLO PDF set~\cite{Butterworth:2015oua}. For Higgs boson production in association with a heavy quark pair, events were simulated  with\progname{MadGraph5\_aMC@NLO} (v.2.2.3 for  \ttH\  and v.2.3.3 for \bbH)~\cite{Alwall:2014hca}, using the CT10 NLO PDF set~\cite{Lai:2010vv} for \ttH\ and the NNPDF23 PDF set~\cite{Ball:2012cx} for \bbH. For the $gg$F, VBF, $VH$, and \bbH\ production mechanisms,\progname{Pythia8}~\cite{Sjostrand:2007gs} was used for the \hfourl\ decay as well as for parton showering, hadronisation and underlying event using the AZNLO parameter set~\cite{STDM-2012-23}.
For showering in the~\ttH~process,~\progname{Herwig++}~\cite{Bahr:2008pv} was used with the UEEE5 parameter set~\cite{Seymour:2013qka}. The Higgs boson production cross-sections and decay branching ratios, as well as their uncertainties, are taken from Refs.~\cite{LHCHiggsCrossSectionWorkingGroup:2011ti,LHCHiggsCrossSectionWorkingGroup:2012vm,Heinemeyer:2013tqa}. This background is approximately 64\% of the total background prediction.

\item[]\textbf{\ZZllll:} The non-resonant SM \ZZllll~ process was simulated using\progname{Sherpa} 2.2.2~\cite{Gleisberg:2008ta,Gleisberg:2008fv,Cascioli:2011va}
for quark-antiquark annihilation, using the NNPDF3.0 NNLO PDF set. The loop-induced $gg$-initiated \zzstar\ production was modelled with %
\textsc{gg2vv}~\cite{Kauer:2015dma} interfaced to\progname{Pythia8}, where \textit{s}-channel $H$ diagrams were omitted to avoid double-counting this contribution, using the CT10 PDFs. The latter process was calculated at LO and receives large QCD corrections at NLO. The cross-section of the sample was therefore multiplied by an NLO/LO $K$-factor of $1.70 \pm 0.15$~\cite{Caola:2015psa}.
This background contributes with  approximately 30\% of the total prediction.
\item[]\textbf{$VVV$, $t\bar t+V$:} The triboson backgrounds $ZZZ$, $WZZ$, and $WWZ$ with four or more leptons originating from the hard scatter were produced with\progname{Sherpa} 2.1.1~\cite{Gleisberg:2008ta,Hoeche:2009rj,Gleisberg:2008fv,Schumann:2007mg,Schonherr:2008av,Cascioli:2011va,Hoeche:2012yf}.
 The all-leptonic \ttZ\ and \ttW\ processes were simulated with \progname{MadGraph5\_aMC@NLO} interfaced to\progname{Pythia8} with the A14 parameter set.
This background is approximately 0.5\% of the total prediction.

\item[]\textbf{Reducible background:} Processes like \zjets, $t\bar{t}$ and $WZ$, produce less than four prompt leptons but can contribute to the selection through jets misidentified as leptons. \zjets~events were modelled using\progname{Sherpa} 2.2.2. The $t\bar{t}$ background was generated with\progname{Powheg-Box} interfaced to\progname{Pythia6}~\cite{Sjostrand:2006za} for parton shower and hadronisation and underlying event. The $WZ$  production was modelled using\progname{Powheg-Box} plus\progname{Pythia8} and the AZNLO parameter set. This background is approximately 6\% of the total prediction.
\end{itemize}

The generation of the simulated samples includes the effect of multiple $pp$ interactions in the same and nearby bunch crossings (pile-up). This was simulated at LO with\progname{Pythia8} using MSTW 2008 PDFs~\cite{Martin:2009iq} and the A2 tune~\cite{ATL-PHYS-PUB-2012-003}. %
The samples were then passed through a simulation of the ATLAS detector~\cite{SOFT-2010-01} based on \textsc{GEANT}4~\cite{Agostinelli:2002hh}.
Weights were applied to the simulated events to correct for the small differences relative to data in the reconstruction, identification, isolation, and impact parameter efficiencies for electrons and muons~\cite{PERF-2016-01,PERF-2015-10}. Furthermore, the lepton momentum scales and resolutions were adjusted to match the data~\cite{ATL-PHYS-PUB-2016-015,PERF-2015-10}.

\subsection{Event selection}
\label{sec:ZZdAnalysis_EventSelection}

All possible combinations of quadruplets are formed by selecting two same-flavour opposite-sign (SFOS) lepton pairs. Each quadruplet must not include more than one stand-alone or calorimeter-tagged muon, and its three leading leptons must have \pt\ (\et) $>$ 20, 15, 10~\gev.\footnote{The \pt requirement is applied to muons while the \et requirement is applied to electrons.} Then a quadruplet per final state is chosen so that the leading pair is defined as the SFOS pair with the mass $m_{12}$ closest to the $Z$ boson mass and the subleading pair is defined as the SFOS pair with the mass $m_{34}$ second closest to the $Z$ boson mass. From this point, the analysis selection proceeds in parallel for the four final states ($4\mu$, $2e2\mu$, $2\mu2e$, $4e$).

For each final state, $m_{12}$ is required to be in the range of $50-106$~\gev, while $m_{34}$ is required to be in the range of $12$--$115$~\gev. A separation of $\Delta R>0.10$ (0.20) is required for all possible pairings of same-flavour (different-flavour) leptons in the quadruplet.  Quadruplets are removed if an alternative same-flavour opposite-sign dilepton mass is less than 5~\gev.  Then the loose calorimeter- and track-based isolation as well as impact parameter requirements explained in Section~\ref{sec:object_selection} are imposed on the leptons. As the four leptons should originate from a common vertex point, a requirement on the $\chi^2$ value of a common-vertex fit is applied, corresponding to a signal efficiency of 99.5\% for all decay channels.
If more than one quadruplet passes the selection, the channel
with the highest expected signal rate is selected, in the order: $4\mu$, $2e2\mu$, $2\mu 2e$ and $4e$.  At this point only one quadruplet remains. In order to improve the four-lepton mass reconstruction, final-state radiation photons arising from any of the lepton candidates in the quadruplet are added to the $4\ell$ system using the same strategy as in Ref.~\cite{Aad:2014eva}. Events are then classified into $2\ell 2\mu$ and $2\ell 2e$ final states. The signal region is defined by the window of  the four-lepton invariant mass $m_{4\ell}\in [115,130]$~\gev.

\subsection{Background estimation}
\label{sec:ZZdAnalysis_BackgroundEstimation}

The dominant background contribution comes from \HZZllll, followed by non-resonant SM $ZZ^*$ production. Triboson processes as well as $t\bar{t}+V$ processes are sources of smaller backgrounds. The background processes described above are estimated from simulation and normalised with the theoretical calculations of their cross-section as described in Section~\ref{sec:ZZdAnalysis_MCsimulation}.

The reducible background  is estimated using data-driven techniques.  Different approaches are followed for the $2\ell 2\mu$ and $2\ell 2e$ final states~\cite{Aad:2014eva}. The procedure to estimate the normalisation of these backgrounds is explained in Ref.~\cite{HIGG-2016-25}. The shapes of the $Z$+jets and $t\bar t$ backgrounds for the $m_{34}$ distribution are taken from simulation and normalised using the inclusive data-driven estimate. For the $WZ$ production, as the background sources are different between the two channels,  this background is included in the data-driven estimate for the $2\ell 2e$ final
state, while it is added from simulation for the $2\ell 2\mu$ final state.

\subsection{Systematic uncertainties}
\label{sec:ZZdAnalysis_SystematicUncertainties}

Imperfect knowledge of the parameters affecting the measurements either from simulated or from data-driven estimates leads to systematic uncertainties which affect the normalisation or the shape of the signal and background samples. Each source of systematic uncertainty is considered to be uncorrelated with other sources. They are listed below.

{\bf Luminosity and pile-up:} The uncertainty in the integrated luminosity is
3.2\%, affecting the overall normalisation of all processes estimated from the
simulation.
It is derived, following a methodology similar to that detailed in Ref.~\cite{DAPR-2011-01}, from a calibration of the luminosity scale using $x-y$ beam-separation scans performed in August 2015 and  May 2016.
The uncertainty associated with the modelling of pile-up arises mainly from
differences between the expected and observed fraction of the visible $pp$
cross-section.

{\bf Lepton-related uncertainties:} Uncertainties associated with leptons arise
from the reconstruction and identification
efficiencies~\cite{PERF-2016-01,PERF-2015-10}, as well as lepton momentum
scales and resolutions~\cite{ATL-PHYS-PUB-2016-015,PERF-2015-10}.
The efficiencies are measured using tag-and-probe techniques on $Z \rightarrow
\ell^+\ell^-$, $J/\psi \rightarrow \ell^+\ell^-$ and $\Upsilon \rightarrow
\mu^+\mu^-$ data and simulated events.
The small differences found are corrected for in the simulation.
The combined effect of all these uncertainties results in an overall
normalisation uncertainty in the signal and background ranging up to 10\%. The dominant uncertainties arise in the reconstruction
and identification of leptons.

{\bf MC background modelling:} Uncertainties in the factorisation and
renormalisation scales, the parton shower, the choice of PDF, and the
hadronisation and underlying-event model affect those backgrounds normalised
with their theory cross-sections.
Uncertainties in the modelling of \HZZllll{} are found to be between 3\% and 9\% depending on
the Higgs boson production mode, while for Standard Model $\qqbar/gg
\rightarrow ZZ^{*}$ processes uncertainties from these sources add in quadrature to 5\%.

{\bf Signal modelling:} Several sources of systematic uncertainty affect the
theoretical modelling of the signal acceptance. Uncertainties originating from
the choice of PDFs, the factorisation and renormalisation scales, and the modelling
of parton shower, hadronisation, and underlying-event account for a total
effect of 9\%~\cite{LHCHiggsCrossSectionWorkingGroup:2011ti,
LHCHiggsCrossSectionWorkingGroup:2012vm}.

{\bf Data-driven estimation of the background:} Uncertainties coming from the data-driven estimation of the background are also considered. They depend on the channel and affect the normalisation of the reducible background~\cite{HIGG-2016-25}.

\subsection{Results}
\label{sec:ZZdAnalysis_Results}

\begin{figure}[!htbp]
	\begin{center}
		\includegraphics[width=0.5\textwidth]{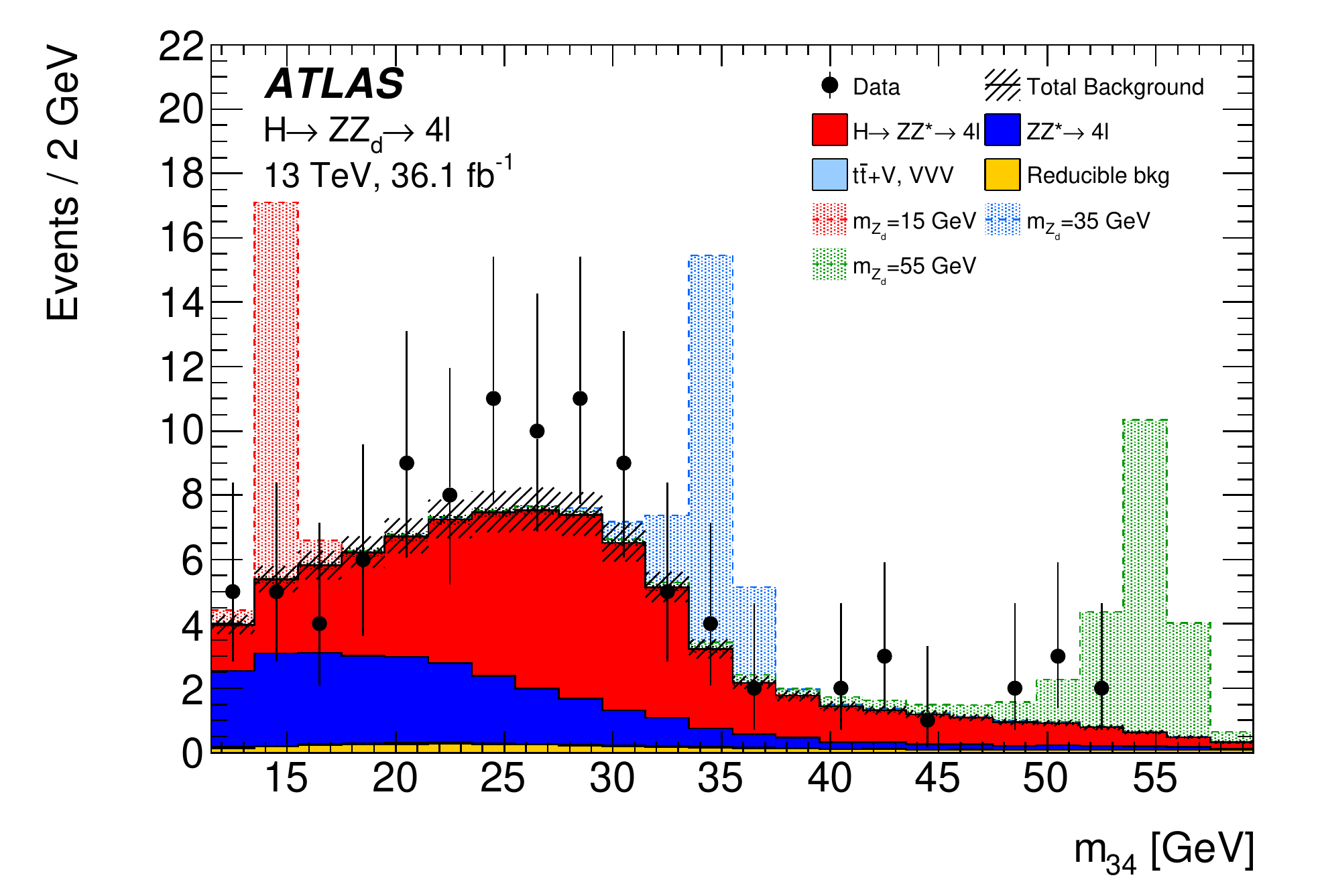}
		\caption{Distribution of $m_{34}$ for data and background events in the mass range $m_{4\ell} \in [115,130]$~\GeV~after the  $H\rightarrow ZX\rightarrow 4\ell$~selection. Three signal points for the \HZZpllll~model are shown. The signal strength corresponds to a branching ratio $\BR(H\rightarrow ZZ_d \rightarrow 4\ell) = \frac{1}{3}\BR(H\rightarrow ZZ^*\rightarrow 4\ell)$ (with $\BR(H\rightarrow ZZ^*\rightarrow 4\ell)$ corresponding to the SM prediction~\cite{LHCHiggsCrossSectionWorkingGroup:2012vm}). The uncertainties include statistical and systematic contributions.  %
			\label{fig:distri_m34}}
	\end{center}
\end{figure}

   The distribution of the  invariant mass of the subleading dilepton pair $m_{34}$ in the selected events in all four final states is shown in Figure~\ref{fig:distri_m34}.  The numbers of events observed in each channel after the event selection, as well as the expected background, are presented in Table~\ref{tab:yield_bkg}. A total of 102 events are observed for an expected background of $86.8 \pm  7.5$. %

 \begin{table}[htbp]
\begin{center}
\scalebox{1.0}{\begin{tabular}{l|c|c|c}
\toprule
 Process &   $2\ell 2\mu$  & $2\ell2e$ & Total \\
\hline
$H\rightarrow ZZ^{*}\rightarrow 4\ell$   & $34.3 \pm 3.6$         & $21.4 \pm 3.0$       &   $55.7 \pm 6.3$ \\
  $ZZ^{*}\rightarrow 4\ell$   			     &  $16.9 \pm 1.2$        & $\hd9.0  \pm 1.1$    &  $25.9 \pm 2.0$ \\
 Reducible background	                   & \hd$2.1  \pm 0.6 $     & \hd$2.7  \pm 0.7$    & \hp$4.8 \pm 1.1$ \\
  $VVV$, $t\bar t+V$                     &  \hd$0.20 \pm 0.05$    & \hd$0.20  \pm 0.04$  &  \hp$0.40 \pm 0.06$\\
\hline
Total  expected	                         &  $53.5 \pm 4.3$  & $33.3 \pm 3.4$ & $86.8 \pm  7.5$ \\
 \hline
  			     Observed &  65 & 37 & 102 \\

\bottomrule

\end{tabular}}

\caption{Expected and observed numbers of events in each channel after the  $H\rightarrow ZX\rightarrow 4\ell$~event selection defined by the mass range $m_{4\ell} \in [115,130]$~\GeV. The uncertainties include MC-statistical and systematic components. \label{tab:yield_bkg}}
\end{center}

\end{table}%

\section{$H\rightarrow XX \rightarrow 4\ell$ ($15~\GeV < m_{X} < 60~\GeV$) 
analysis}
\label{sec:ZdZdHighMassAnalysis}

\subsection{Monte Carlo simulation}
\label{sec:ZdZdHighMassAnalysis_MCsimulation}

The signal process $H \rightarrow Z_dZ_d \rightarrow 4\ell$ was generated using 
the same model and event generator as in the $H \rightarrow ZZ_d \rightarrow 4\ell$ 
analysis. The model parameters $\epsilon$ and $\kappa$ were adjusted so that only \HZpZpllll~decays were generated ($\kappa \gg \epsilon$).
Contributions from final states with $\tau$ leptons
are not considered. The mass of the $Z_d$ boson was varied for different signal 
hypotheses in the range $15~\GeV < m_{Z_d} < 60~\GeV$ for $H \rightarrow Z_dZ_d 
\rightarrow 4\ell$, in steps of 5~\GeV. The same signal normalisation was also 
applied (normalising to the SM Higgs boson production cross-section and 
$\BR(H\rightarrow ZZ^* \rightarrow 4\ell)$). 

For the signal process  $H \rightarrow aa \rightarrow 4\mu$, the Higgs boson 
production was generated with the\progname{Powheg-Box} v2 event 
generator~\cite{Nason:2004rx,Frixione:2007vw,Alioli:2010xd} using the CT10 NLO 
PDFs~\cite{Lai:2010vv}. Then this SM Higgs boson is replaced by a neutral 
scalar Higgs boson from the 2HDM+S~\cite{a0} model. \progname{Pythia8} was used for 
the showering, hadronisation and underlying-event simulation in events 
generated where the Higgs boson decays to two pseudoscalar bosons that 
subsequently decay to pairs of muons, 
$H\rightarrow aa\rightarrow 4\mu$. The $a$-boson decay was 
done in the narrow-width approximation and the coupling to the muons is made 
to be that of a pseudoscalar. The mass of the $a$-boson was varied for five 
different signal hypotheses in the range $15~\GeV < m_{a} < 60~\GeV$.

The background processes considered in this analysis are similar to those 
featured in the $H\rightarrow ZZ_{d}\rightarrow 4\ell$ analysis described in 
Section~\ref{sec:ZZdAnalysis}, however all the background processes are now 
reducible by vetoing on lepton pairs consistent with a $Z$ boson (see 
section~\ref{sec:ZdZdHighMassAnalysis_EventSelection} for more details). The 
background processes, in order of 
decreasing importance, are as follows:

\begin{itemize}
	\item[] \textbf{\HZZllll:} %
	The modelling of this process is the same as for the $H\rightarrow ZZ_{d} 
	\rightarrow 4\ell$ analysis, as described in 
	Section~\ref{sec:ZZdAnalysis_BackgroundEstimation}. 
	This background is approximately 63\% of the total background prediction 
	for this analysis.
	\item[] \textbf{\ZZllll:} The dominant $q\bar{q}$ production mechanism was modelled by\progname{Powheg-Box} interfaced to\progname{Pythia8}. The $gg$-initiated production mechanism was modelled in the same way as described for the  $H\rightarrow ZZ_{d} \rightarrow 4\ell$ analysis.
	Both production mechanisms use the CT10 PDFs.
	This background is approximately 19\% of the total background prediction.%
	\item[] \textbf{VVV/VBS:} Higher-order electroweak	
	processes (with cross-sections proportional to $\alpha^6$ at leading order) 
	include triboson production and vector-boson scattering, which lead to four 
	leptons in the final state, with two additional particles (quarks, 
	neutrinos, or electrons and muons). These processes were modelled by 
	\textsc{Sherpa}~2.1.1 with the CT10 PDFs. Higgs boson production through VBF is 
	subtracted from the estimates obtained with this event generator, in order to 
	avoid double-counting this process in final background estimates.
	This background is approximately 17\% of the total background prediction.
	\item[] \textbf{$Z+(t\bar{t}/J/\psi/\Upsilon) \rightarrow 4\ell$:} This background 
  process corresponds to the production 
	of a $Z$ boson, in association with either a quarkonium state ($b\bar{b}$ 
	or $c\bar{c}$) that decays to leptons, or $t\bar{t}$ production with 
	leptonic decays of the prompt $W$ bosons. The  processes involving 
	quarkonia were modelled using\progname{Pythia8} with the NNPDF 2.3 PDFs~\cite{Ball:2012cx}. The 
	\ttZ\ process was modelled at leading order with 
	\textsc{MadGraph}5~\cite{Alwall:2011uj} interfaced to\progname{Pythia8}, 
	using the NNPDF 2.3 PDFs~\cite{Ball:2012cx} and the A14 
	tune~\cite{cite:autwo}.
	These backgrounds are approximately 1\% of the total background prediction.
	\item[] \textbf{Other reducible backgrounds:} %
	The same three SM processes of diboson $WZ$ production, production of $t\bar{t}$ which decay 
	semileptonically or fully leptonically and single $Z$ production with 
	associated jets are considered as in the $H\rightarrow ZZ_{d} \rightarrow 
	4\ell$ analysis, with the same modelling for the first two. The last one 
	however was modelled by\progname{Powheg-Box} interfaced to\progname{Pythia8} and 
	the CTEQ6L1 PDFs~\cite{CTEQ6L1}. The contribution from these processes is estimated to be 
	negligible. 
	
\end{itemize}

\subsection{Event selection}
\label{sec:ZdZdHighMassAnalysis_EventSelection}

After forming the possible SFOS quadruplets in each event (as described in 
Section~\ref{sec:ZZdAnalysis_EventSelection}), the following selections are 
applied to the quadruplets: each quadruplet must not include more than one 
stand-alone or calorimeter-tagged muon, and the three leading leptons in the 
quadruplet must have \pt\ $>$ 20, 15, 10~\gev. Additionally, the leptons 
in the quadruplet must be matched to at least one of the triggers used to 
select the event at trigger-level, and a separation of $\Delta R > 0.1 (0.2)$ 
is required for all same-flavour (different-flavour) leptons in the quadruplet. The 
event is discarded if no quadruplets remain. From any quadruplets remaining, a 
single quadruplet is selected as the one with the smallest dilepton 
invariant mass difference $\delta m = |m_{12}-m_{34}|$. This procedure for 
selecting a single quadruplet can result in the incorrectly paired quadruplet 
being selected in $4e$ or $4\mu$ signal events. The fraction of signal events 
where this occurs was estimated using the $\Zp\Zp$ MC samples to be 
approximately 2\% (1\%) in the $4e$ ($4\mu$) channel for $m_X = 15~\GeV$, 
rising to 8\% (5\%) at $m_X = 60~\GeV$.  Events are classified 
into three channels according to the flavours of the leptons in the selected 
quadruplet: $4e, 2e2\mu$, and $4\mu$ (no distinction is made between $2e2\mu$ 
and $2\mu2e$ permutations).

The remaining selections are applied to the selected quadruplet of the event, 
with the event discarded if any selection fails: the four-lepton invariant mass 
must be in the range $115 < m_{4\ell} < 130~\GeV$, to select events consistent 
with a 125~\GeV{} Higgs boson. The ratio of the secondary dilepton's mass to the primary 
dilepton's mass ($m_{34}/m_{12}$) must be greater than 0.85, which selects events 
where the dilepton masses are similar. Neither of the dilepton invariant masses 
is allowed to be in a mass range around the $J/\Psi$ or $\Upsilon$ resonance masses (see 
Table~\ref{tab:eventselection}), with this requirement also 
applied to the alternative-pairing dilepton invariant masses ($m_{14}$ and 
$m_{32}$) for events with a $4e$ or $4\mu$ selected quadruplet. Finally, the 
dilepton invariant masses are required to be in the range $10 < m_{12,34} < 
64~\GeV$ and in the case of $4e$ and $4\mu$ quadruplets the alternative-pairing 
dilepton masses must be in the range $5 < m_{14,32} < 75~\GeV$. These last two 
selections suppress backgrounds that contain a $Z$ boson, and are referred to 
as the \emph{Z Veto} in the following.

\subsection{Background estimation}
\label{sec:ZdZdHighMassAnalysis_BackgroundEstimation}

All background estimates for this analysis rely on using MC simulations, 
and are validated in regions that are orthogonal to the signal event 
selection described in the previous section.

\begin{figure}[!htbp]
	\begin{center}
		\subfloat[VR1]{\includegraphics[width=0.49\textwidth]{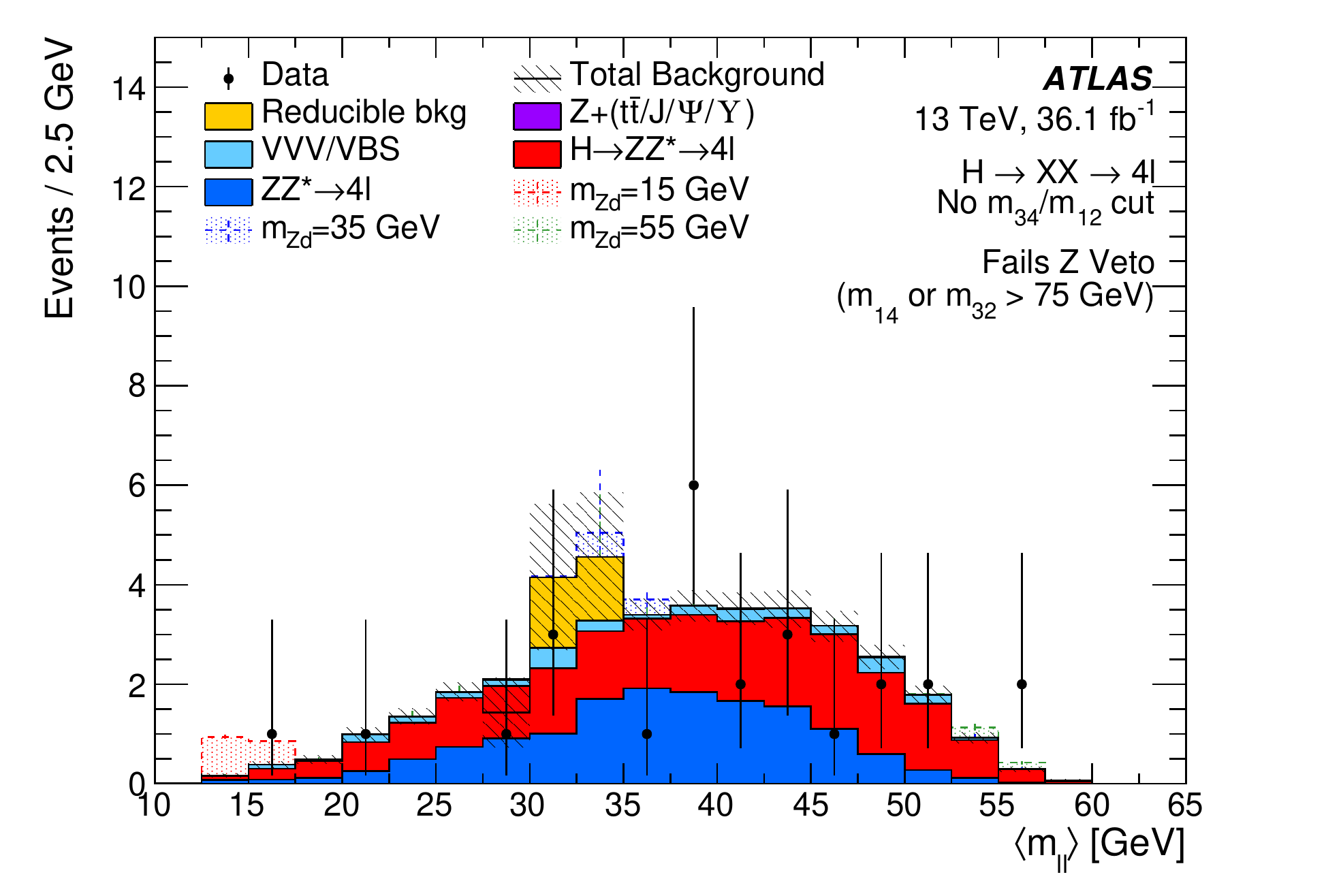}}
		\subfloat[VR2]{\includegraphics[width=0.49\textwidth]{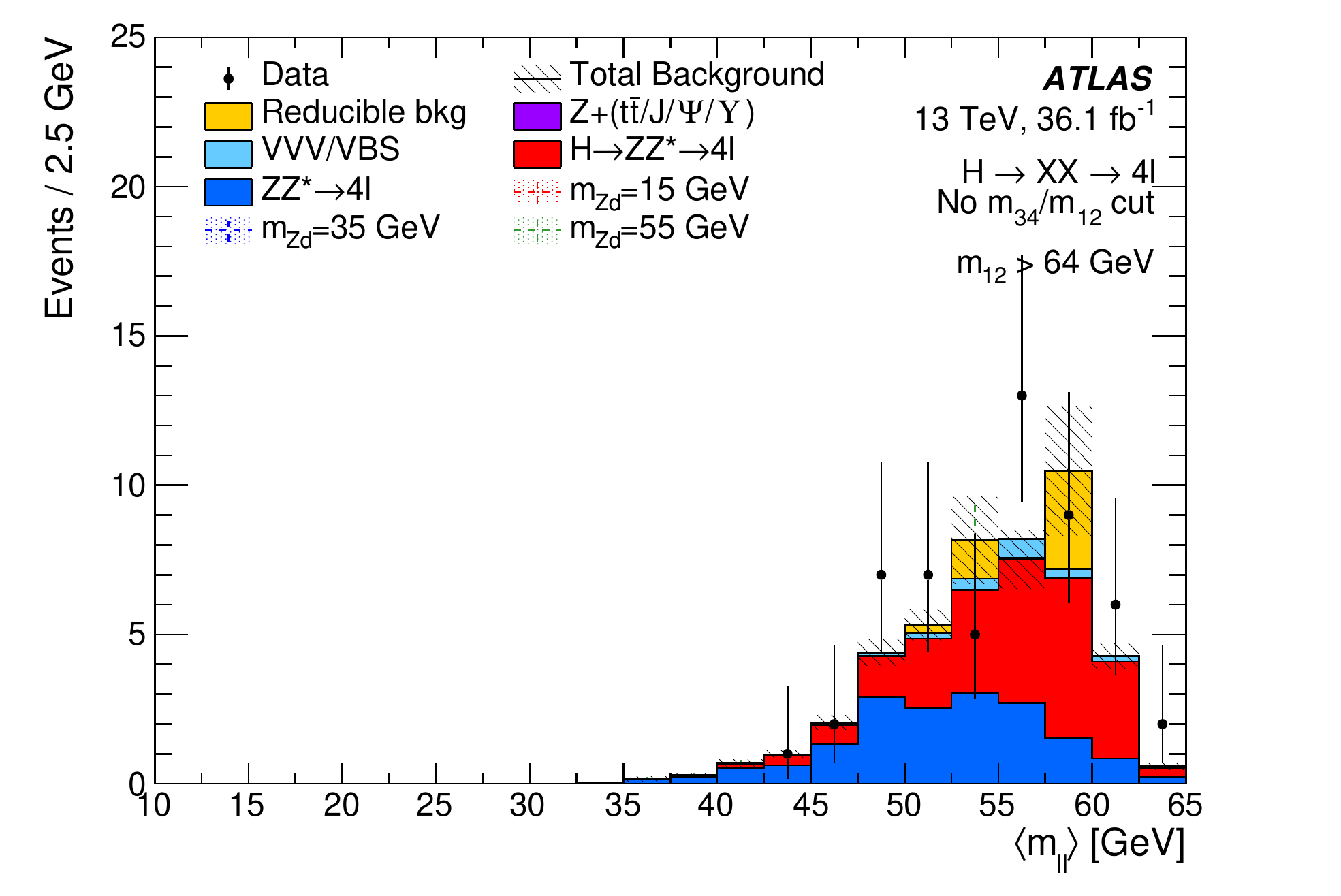}} \\
		\subfloat[VR3]{\includegraphics[width=0.49\textwidth]{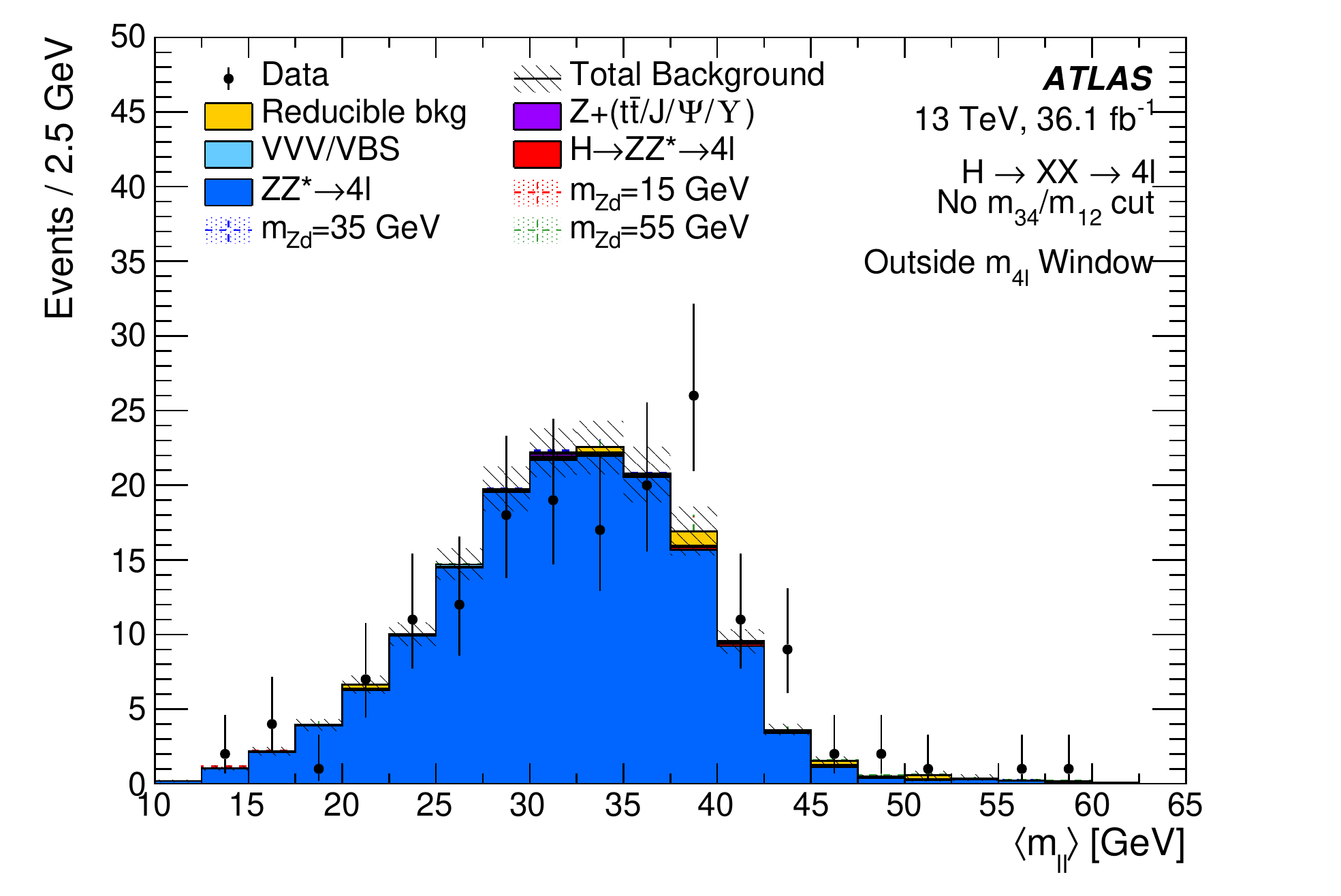}}
        \caption{Distributions of $\langle m_{\ell\ell}\rangle= \frac{1}{2}(m_{12}+m_{34})$ in three background validation regions of the $H\rightarrow XX \rightarrow 4\ell (15 < m_{X} < 60~\GeV)$ analysis: (a) events failing the \emph{Z Veto} ($4e$ or $4\mu$ events where $m_{14}$ or $m_{32} > 75~\GeV$), (b) events where $m_{12}>64~\GeV$, (c) events outside of the $115 < m_{4\ell} < 130~\GeV$ window. In all cases the $m_{34}/m_{12}$ requirement is removed to increase the number of events. The (negligible) contamination by the signal in these validation regions is shown for three mass hypotheses of the vector-boson benchmark model: the signal strength corresponds to a branching ratio $\BR(H\rightarrow Z_d Z_d \rightarrow 4\ell) = \frac{1}{10}\BR(H\rightarrow ZZ^*\rightarrow 4\ell)$ (with $\BR(H\rightarrow ZZ^*\rightarrow 4\ell)$ corresponding to the SM prediction~\cite{LHCHiggsCrossSectionWorkingGroup:2012vm}).
			\label{fig:ZdZdHighMassAnalysis_validation}}
	\end{center}
\end{figure}

The two main background processes ($\HZZllll$ and $\ZZllll$) are validated by 
comparison of the background prediction to data in three validation regions 
that are orthogonal to the signal region. The first validation region 
(VR1) is defined by reversing part of the \emph{Z Veto} requirement: VR1 
requires $m_{14}$ or $m_{32}$ to be greater than $75~\GeV$. The second 
validation region (VR2) instead requires that $m_{12}$ is greater than 
$64~\GeV$. These two regions primarily validate the $\HZZllll$ prediction. The 
third validation region (VR3) reverses the requirement on the four-lepton invariant mass window, 
i.e.\ requires $m_{4\ell}<115~\GeV$ or $m_{4\ell} > 130~\GeV$. In 
all three validation regions the $m_{34}/m_{12} > 0.85$ requirement is removed 
in order to increase the number of data events. Distributions of the average dilepton 
mass are shown for the three validation regions in 
Figure~\ref{fig:ZdZdHighMassAnalysis_validation}.

The background estimates for the signal region are given in 
Table~\ref{tab:ZdZdHighMassAnalysis_yields}, and include all systematic 
uncertainties described in 
Section~\ref{sec:ZdZdHighMassAnalysis_SystematicUncertainties}.

\subsection{Systematic uncertainties}
\label{sec:ZdZdHighMassAnalysis_SystematicUncertainties}

The systematic uncertainties in the signal and background modelling are the 
same as those described in 
Section~\ref{sec:ZZdAnalysis_SystematicUncertainties} (excluding the 
data-driven background uncertainties, which are not applicable to this search).
It should be noted that fewer than four background events are predicted in the 
signal region for the $H \rightarrow XX \rightarrow 4 \ell$ analysis, and 
therefore the dominant uncertainty in the background prediction is the 
statistical uncertainty.

\subsection{Results}
\label{sec:ZdZdHighMassAnalysis_Results}

\begin{figure}[!htbp]
	\begin{center}
		\subfloat[Signal region $\langle m_{\ell\ell}\rangle$ 
		distribution]{\includegraphics[width=0.49\textwidth]{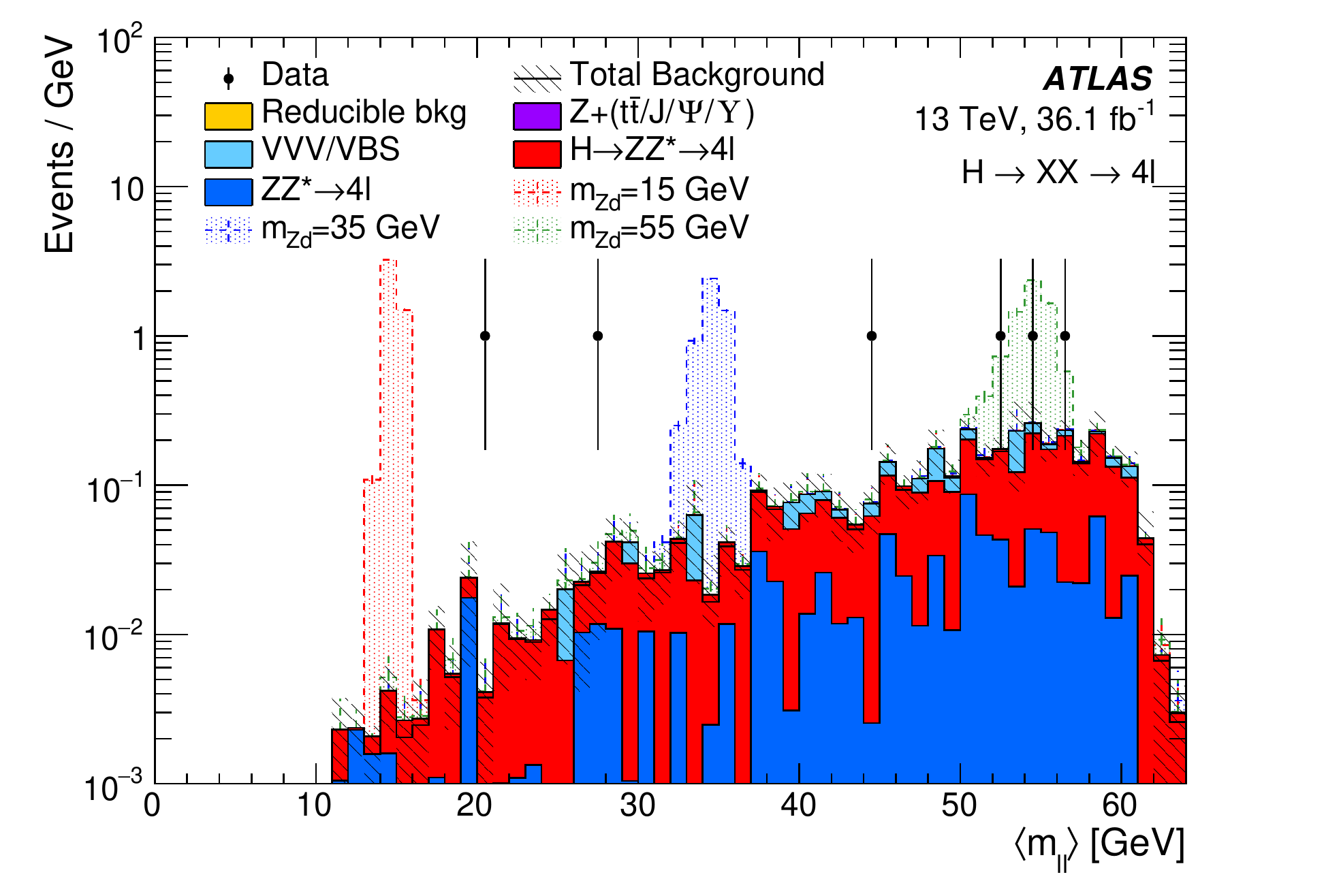}}
		\subfloat[$m_{34}$ vs $m_{12}$ 
		distribution\label{fig:resultsDataScatterHighMass}]{\includegraphics[width=0.49\textwidth]{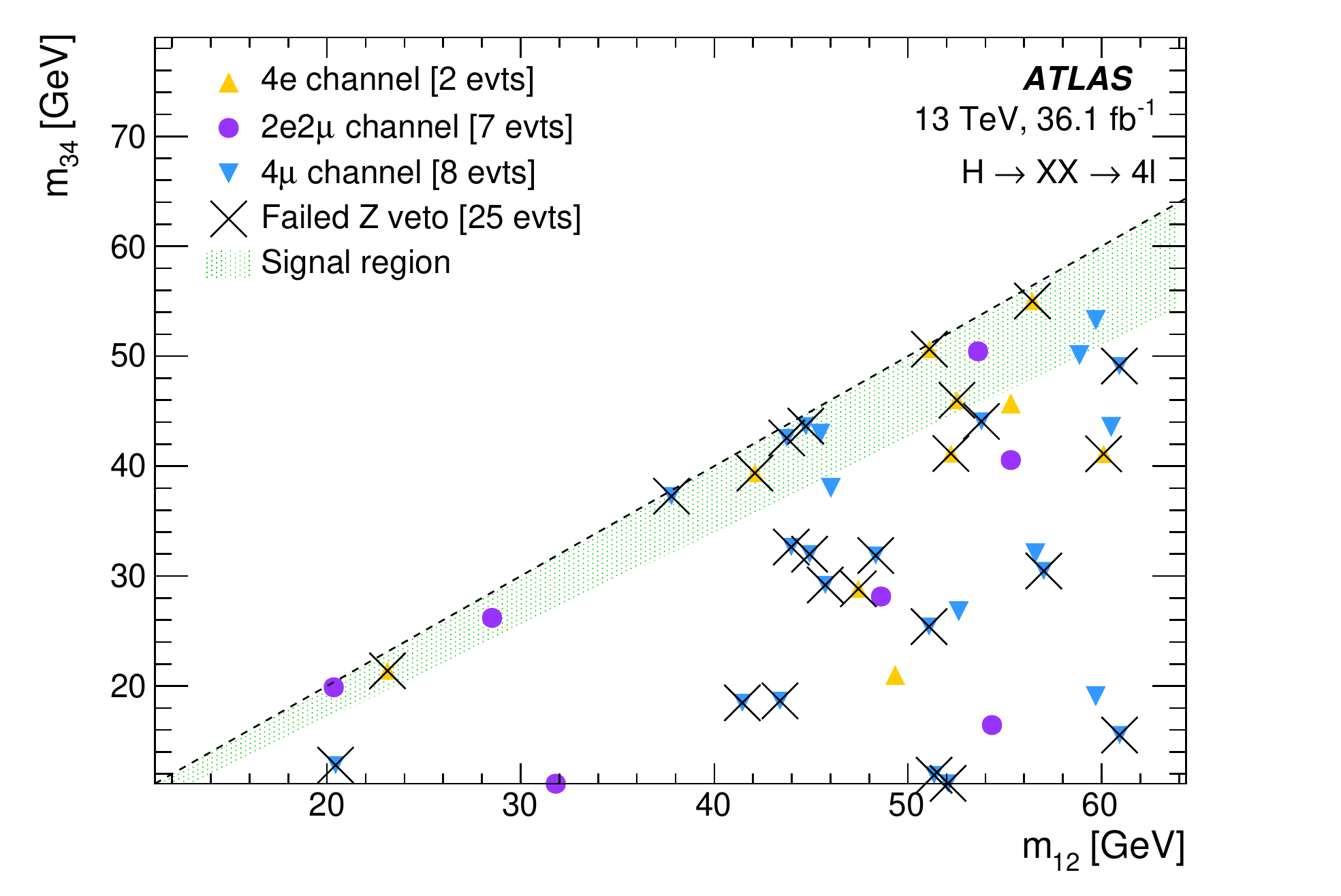}}
		
    \caption{ Distribution of (a) $\langle m_{\ell\ell}\rangle= 
    \frac{1}{2}(m_{12}+m_{34})$ and (b) $m_{34}$ vs $m_{12}$, for events 
    selected in the $H\rightarrow XX \rightarrow 4\ell~(15 < m_{X} < 60~\GeV)$ 
    analysis. The example signal distributions in (a) correspond to the 
    expected yield normalized with $\sigma(pp\rightarrow\HZpZpllll) = 
    \frac{1}{10}\sigma_{\text{SM}}(pp\rightarrow\HZZllll)$. The crossed-through 
    points in (b) fail the \emph{Z Veto}. The events outside the (shaded green) 
    signal region in figure (b) are events that fail the $m_{34}/m_{12} > 0.85$ 
    requirement. The diagonal dashed line marks where $m_{12}=m_{34}$, and in 
    this range of dilepton masses all events will have $m_{34}<m_{12}$.
			\label{fig:ZdZdHighMassAnalysis_results}}
	\end{center}
\end{figure}

The distributions of $\langle m_{\ell\ell}\rangle = \frac{1}{2}(m_{12}+m_{34})$ 
for the events selected in this analysis are shown in 
Figure~\ref{fig:ZdZdHighMassAnalysis_results}, and the total yields presented in Table~\ref{tab:ZdZdHighMassAnalysis_yields}:
six events are observed for a 
prediction of $3.9 \pm 0.3$ events in the high-mass selection. 

\begin{table}[htbp!]
	\begin{center}
		\begin{small}
			\begin{tabular*}{0.4\textwidth}{@{\extracolsep{\fill}}lc}
				\toprule
				\textbf{Process} & Yield \\
				\hline
				$\ZZllll$          & $0.8\pm0.1$ \\
				$\HZZllll$         & $2.6\pm0.3$   \\
				VVV/VBS               & $0.51\pm0.18$ \\
				$Z + (t\bar{t}/J/\Psi)\rightarrow 4\ell$   & $0.004\pm0.004$ \\
				Other Reducible Background & Negligible \\
				\hline
				Total              & $3.9\pm0.3$   \\
				\hline
				Data               & 6             \\
				\bottomrule
			\end{tabular*}
			\caption{\label{tab:ZdZdHighMassAnalysis_yields}Expected event 
			yields of the SM background processes and observed data in the 
			$H\rightarrow XX \rightarrow 4\ell$~($15~\GeV < m_{X} < 60~\GeV$) 
			selection. The uncertainties include MC-statistical and systematic 
			components.
			}
		\end{small}
	\end{center}
\end{table}

The biggest deviation from the Standard Model expectation is from a single 
event at $\langle m_{\ell\ell}\rangle\approx 20~\GeV$, with a local 
significance of $3.2\,\sigma$. The corresponding global significance is 
approximately $1.9\,\sigma$, estimated using an approximation~\cite{Gross:2010qma} 
for the tail probability of the profile-likelihood-ratio test statistic. The 
significances are calculated using a Gaussian signal model with normalisation, 
yield, and standard deviation determined by interpolation between the 
corresponding fits to simulated signal samples (5~\GeV~intervals in $Z_d$ 
mass). This statistical model, where the signal spans several bins of the 
$\langle m_{\ell\ell}\rangle$ distribution, means that statistical fluctuations 
in the background estimate do not significantly impact the calculation of 
significance.

The $m_{34}$ versus $m_{12}$ distribution of the selected events is shown in 
Figure~\ref{fig:resultsDataScatterHighMass}. In this figure, the 
crossed-through markers correspond to the events that fail the \emph{Z Veto}, 
which required the alternative-pairing masses $m_{32}$ and $m_{14}$ 
(relevant only to the $4e$ and $4\mu$ channels) to be less than $75~\GeV$. This 
requirement has a significant impact on the signal efficiency (up to $\approx 40\%$ 
loss) for $m_X$ just above $15~\GeV$, but is applied in this analysis to 
mitigate any small contributions from SM processes involving $Z$~boson 
production with large cross-sections. The 25 events that fail this veto are 
shown in validation region VR1 in 
Figure~\ref{fig:ZdZdHighMassAnalysis_validation}, where there is a total 
background prediction of $38 \pm 3$ events.

\section{$H\rightarrow XX \rightarrow 4\mu$ ($1~\GeV < m_{X} < 15~\GeV$) 
analysis}
\label{sec:ZdZdLowMassAnalysis}
\subsection{Monte Carlo simulation}
\label{sec:ZdZdLowMassAnalysis_MCsimulation}

The generation of the signal processes $H \rightarrow Z_dZ_d \rightarrow 4\ell$ and $H \rightarrow aa \rightarrow 4\mu$ follows the prescription described in Section~\ref{sec:ZdZdHighMassAnalysis_MCsimulation}.
Four samples were generated with $Z_d$ masses of 1~\GeV, 2~\GeV, 5~\GeV~and 10~\GeV, while the mass of the $a$-boson was varied for 10 different signal hypotheses in the range $0.5~\GeV \leq m_{a} \leq 15~\GeV$.

The background processes considered in this analysis are described in the following:

\begin{itemize}
  \item[] \textbf{\HZZllll:} The modelling of this process is the same as for the $H\rightarrow ZZ_{d} \rightarrow 4\ell$ analysis, described in Section~\ref{sec:ZZdAnalysis_MCsimulation}.

  \item[] \textbf{\ZZllll:} This process was simulated with \textsc{Sherpa}~2.1.1 due to an implicit particle-level requirement on the mass of the $Z^{*}$ in the \textsc{Powheg-Box} MC sample used for the high-mass selection described in Section~\ref{sec:ZdZdHighMassAnalysis_MCsimulation}.
    The $gg$-initiated production mechanism was modelled in the same way as for the  $H\rightarrow ZZ_{d} \rightarrow 4\ell$ analysis, see Section~\ref{sec:ZZdAnalysis_MCsimulation}.
Both production mechanisms are estimated using the CT10 PDFs.

  \item[] \textbf{VVV/VBS:} The modelling of this process is described in Section~\ref{sec:ZdZdHighMassAnalysis_MCsimulation}.

  \end{itemize}

\subsection{Event selection}
\label{sec:ZdZdLowMassAnalysis_EventSelection}

In this search, only events with at least four muons are considered.
Similarly to the searches described above, the selected muons are combined into $4\mu$ quadruplets in all possible permutations of pairs of opposite-sign dimuons.
In the case of having more than four muons, the different quadruplets that can be formed are all considered. Of the muons in each quadruplet, at least three must have $\pt > 10$~\GeV, at least two must have $\pt > 15$~\GeV, and at least one must have $\pt > 20$~\GeV, and there cannot be more than one stand-alone or calorimeter-tagged muon.

The quadruplet selection closely follows the selection described in Section~\ref{sec:ZdZdHighMassAnalysis_EventSelection}.
Nonetheless, low-mass bosons are more boosted and muons less separated.
For this reason, to keep signal efficiencies high, no $\Delta R$ requirement is applied to the muons of the quadruplets.
If more than one quadruplet survives this selection, the one with the smallest $\Delta m_{\ell\ell}$ is selected.

Similarly to the $H\rightarrow XX \rightarrow 4\ell$ ($15 < m_{X} < 60~\GeV$) analysis, a set of requirements are applied to the quadruplet invariant masses as well as to the masses of the different muon pairings in the quadruplet.
The quadruplet invariant mass must satisfy $120~\GeV < m_{4\ell} < 130~\GeV$.
This window is tighter than the selections described in Sections~\ref{sec:ZZdAnalysis_EventSelection} and~\ref{sec:ZdZdHighMassAnalysis_EventSelection} because muons have smaller radiative losses than electrons.
The dilepton masses must be in the range $0.88~\GeV < m_{12,34} < 20~\GeV$.
No restriction is applied to the alternative-pairing dilepton masses because more than one third of signal events in this corner of the phase space contains an alternative-pairing dilepton mass that satisfies $75~\GeV < m_{14,32} < 125~\GeV$, and would therefore be lost if this selection was applied.

\subsection{Background estimation}
\label{sec:ZdZdLowMassAnalysis_BackgroundEstimation}

The main background contributions for this search come from the \ZZllll{} and \HZZllll{} processes, as for the $H\rightarrow XX \rightarrow 4\ell$ ($15 < m_{X} < 60~\GeV$) case described in Section~\ref{sec:ZdZdHighMassAnalysis_BackgroundEstimation}.
These backgrounds, suppressed by the requirements on the lepton invariant mass, account for 30\% each of the total background.
Smaller background contributions come from higher-order
electroweak processes (with cross-sections proportional to $\alpha^6$ at 
leading order) and account for approximately 19\% or the total background.
Finally, events with multiple heavy flavour (bottom or charm) quark decays can also contribute to the total background yield.
A leading part of this contribution comes from double semileptonic decays, where the $b$-quark decays to a muon and a $c$-quark which further decays into another muon and light hadrons.
Resonances produced in the heavy flavour quark decay chain (i.e.\ $\omega$, $\rho$, $\phi$, $\jpsi$) that result in pairs of muons also become an important contribution of this background.
Events with four heavy flavour quarks may pass the signal region requirements if each bottom or charm quark decays semileptonically.

The estimation method for the heavy flavour background was developed using fully data-driven inputs and inspired by a previous analysis from CMS~\cite{CMSa0Limitmumumumu}.
Using data control samples, the background is modelled as a two-dimensional template in the plane of the invariant masses of the two dimuons.
This template is constructed from the Cartesian product of two one-dimensional dimuon invariant mass spectra assuming that each muon pair is independent of the other.
The one-dimensional templates are derived in control regions with three muons passing the same quality and isolation requirements used elsewhere in the analysis.
The {high-\pt} selection requires three muons, with a muon pair with $\pt > 20$~\GeV~and $\pt > 10$~\GeV~matched to the dimuon trigger, and an additional muon with $\pt > 5$~\GeV.
The {low-\pt} selection also requires three muons, with a pair of muons, each with $\pt > 5$~\GeV, and an additional muon with $\pt > 25$~\GeV~matched to the single-muon trigger.
The choice of \pt{} thresholds for the templates was carefully studied and it is required that the first (second) dimuon pair always passes the high-\pt~(low-\pt) selection.
Estimations with alternative \pt{} threshold selections were found to be compatible with the current prescription.
The Higgs boson mass requirement described in Section~\ref{sec:evtsel} introduces a correlation between the dimuon pairs and therefore a correction to the two-dimensional template is necessary.
This correction is extracted from data using a sample enriched in events with heavy flavour quarks, with inverted isolation and vertex requirements.
The final template covers the full $m_{34}$ vs $m_{12}$ plane, including the signal region (defined by the condition $m_{34} / m_{12} > 0.85$).
The normalisation of the template is computed in the region with $m_{34} / m_{12} < 0.85$, and its effect propagated to the signal region.
The heavy flavour processes are negligible in the high-mass region, while they account for 22\% of the total prediction in the low-mass region.

\begin{figure}[!htbp]
	\begin{center}
		\includegraphics[width=0.49\textwidth]{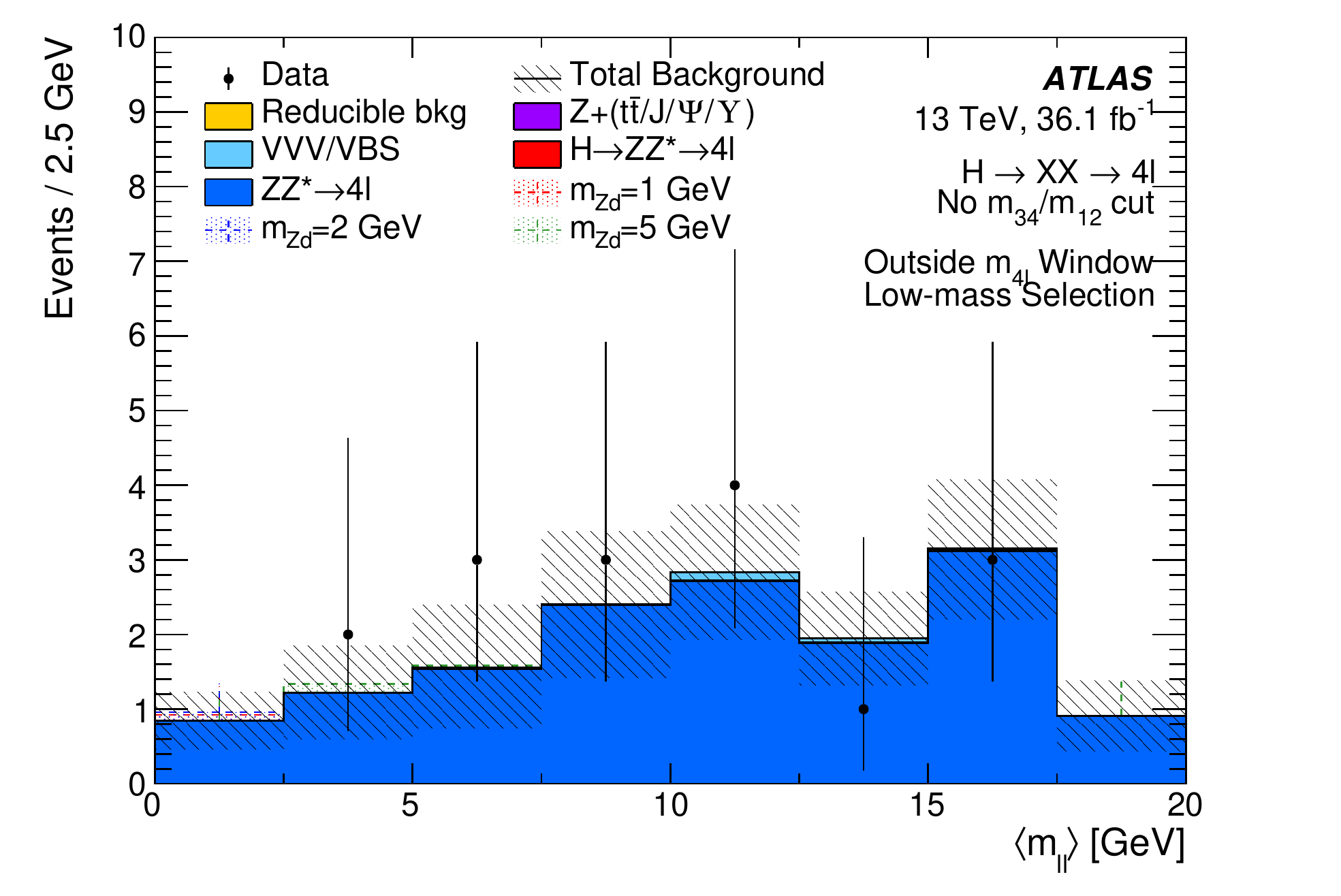}
		\caption{Distribution of $\langle m_{\ell\ell}\rangle= \frac{1}{2}(m_{12}+m_{34})$ in the validation region. The (negligible) contamination by the signal in these validation regions is shown for three mass hypotheses of the vector-boson benchmark model: the signal strength corresponds to a branching ratio $\BR(H\rightarrow Z_d Z_d \rightarrow 4\ell) = \frac{1}{10}\BR(H\rightarrow ZZ^*\rightarrow 4\ell)$ (with $\BR(H\rightarrow ZZ^*\rightarrow 4\ell)$ corresponding to the SM prediction~\cite{LHCHiggsCrossSectionWorkingGroup:2012vm}).
			\label{fig:ZdZdLowMassAnalysis_validation}}
	\end{center}
\end{figure}

The modelling of the most important background processes (\ZZllll{} and \HZZllll{}) was validated in VR1, VR2 and VR3, defined in Section~\ref{sec:ZdZdHighMassAnalysis_BackgroundEstimation} for the $H\rightarrow XX \rightarrow 4\ell$ ($15 < m_{X} < 60~\GeV$) selection.
The \ZZllll{} process can also be validated by comparing the background prediction to the data in a validation region (VR4) that is orthogonal to this signal region.
This validation region is defined by reversing the four-lepton invariant mass window requirement, i.e.\ $m_{4\ell}<120~\GeV$ or $m_{4\ell} > 130~\GeV$.
The average dilepton mass distribution for this region is shown in Figure~\ref{fig:ZdZdLowMassAnalysis_validation}.

\subsection{Systematic uncertainties}
\label{sec:ZdZdLowMassAnalysis_Results}

In addition to the systematic uncertainties described in Section~\ref{sec:ZdZdHighMassAnalysis_SystematicUncertainties}, this analysis includes additional uncertainties for its data-driven background estimate.

Several sources of uncertainty are considered for the heavy flavour background data-driven estimation. Uncertainties in the shape of the one-dimensional templates are propagated to the two-dimensional template to account for shape variations in the dimuon invariant mass spectra.
Different parameterisations of the Higgs boson mass requirement are also considered for modelling the effect of this condition on the shape of the distribution in the $(m_{12}, m_{34})$ plane.
The previous two systematic uncertainties affect the shape of the two-dimensional plane, which propagates to a 63\% effect on the yield of the heavy flavour background in the low-mass search signal region.
Finally, the statistical uncertainty in the normalisation of the template in the signal region ($m_{34}/m_{12} < 0.85$ region of the two-dimensional plane) is also propagated to the signal region to account for fluctuations in the final heavy flavour background yields and has an effect of 13\%.
Uncertainties from the different sources are added in quadrature, and the total amounts to 65\% for this background source.

\subsection{Results}
\label{sec:ZdZdLowMassAnalysis_Results}

\begin{figure}[!htbp]
	\begin{center}
		\subfloat[Signal region $\langle m_{\ell\ell}\rangle$ distribution\label{fig:resultsDataAvgMLowMass}]{\includegraphics[width=0.49\textwidth]{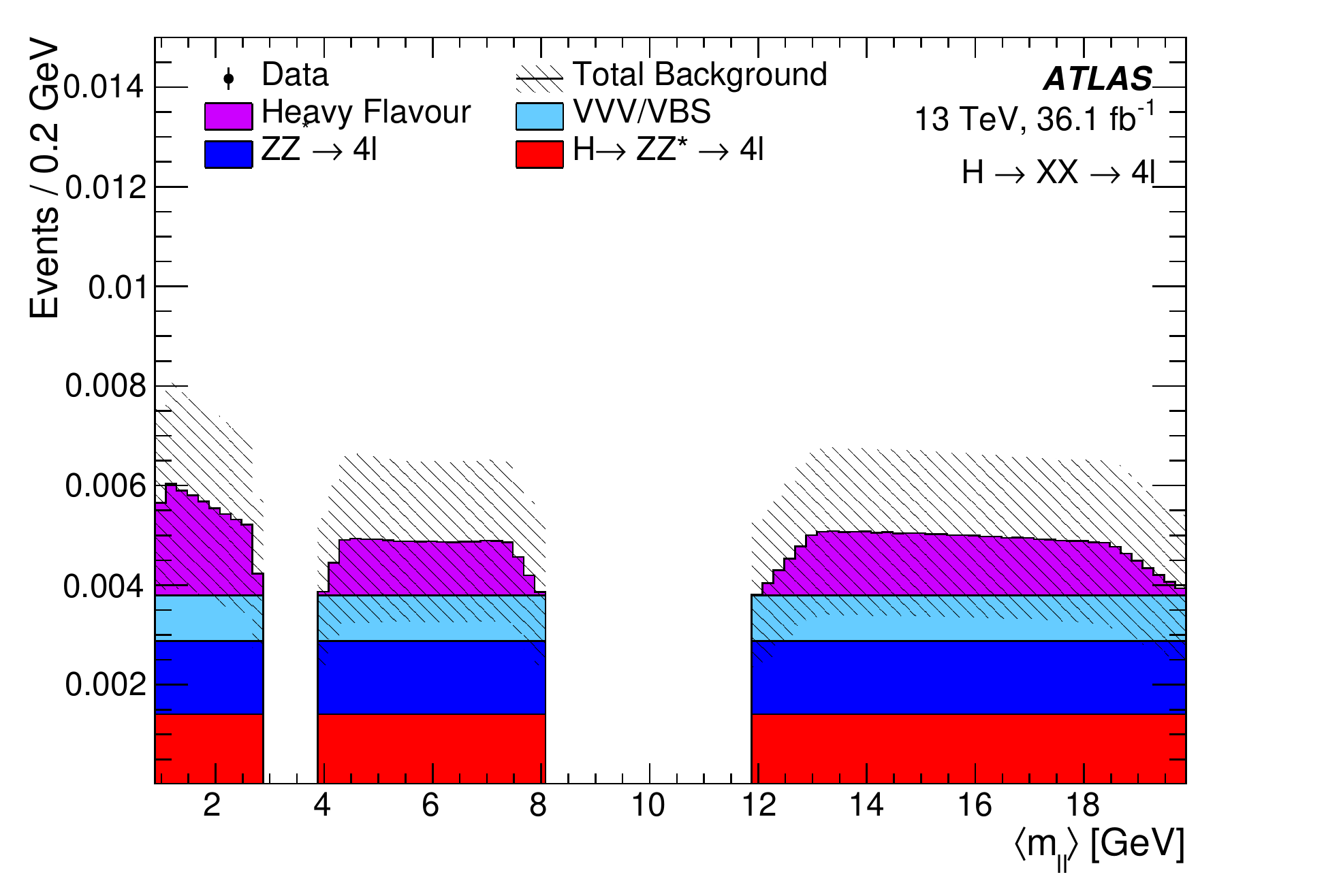}}
		\subfloat[$m_{34}$ vs $m_{12}$ distribution\label{fig:resultsDataScatterLowMass}]{\includegraphics[width=0.49\textwidth]{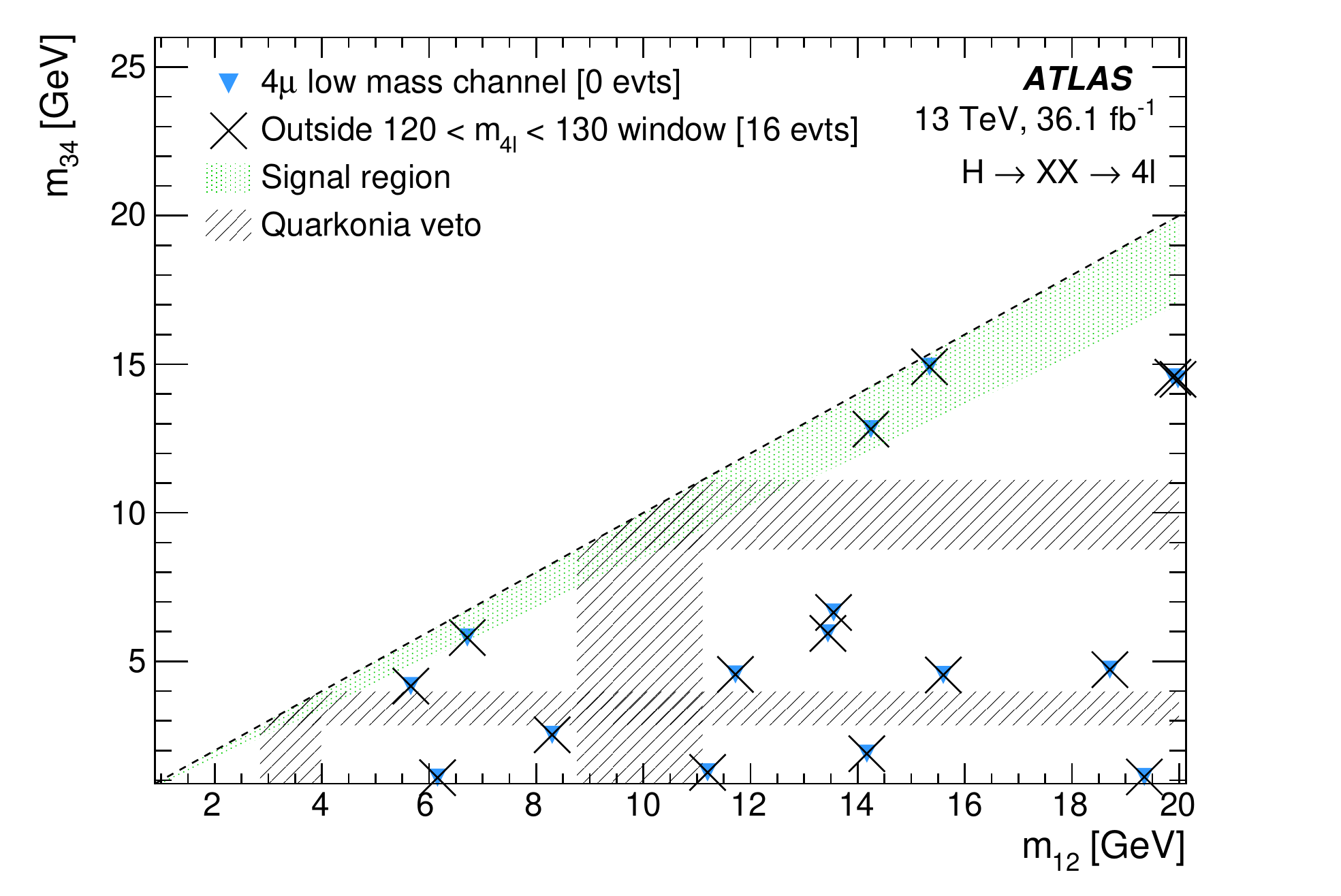}}
		
		\caption{ Distribution of (a) $\langle m_{\ell\ell}\rangle= \frac{1}{2}(m_{12}+m_{34})$ and (b) $m_{34}$ vs $m_{12}$, for events selected in the $H\rightarrow XX \rightarrow 4\mu (1 < m_{X} < 15~\GeV)$ analysis. The crossed-through points in figure (b) correspond to events that are outside the $m_{4\ell}$ mass window of $120~\GeV < m_{4\ell} < 130~\GeV$. The events outside the (shaded green) signal region are events that fail the $m_{34}/m_{12} > 0.85$ requirement.
			\label{fig:ZdZdLowMassAnalysis_results}}
	\end{center}
\end{figure}

The $\langle m_{\ell\ell}\rangle$ distribution for the selected events is shown in Figure~\ref{fig:resultsDataAvgMLowMass}.
Table~\ref{tab:ZdZdLowMassAnalysis_yields} shows the resulting yields and uncertainties for this analysis: no events are observed to pass the selection, for a total background prediction of $0.4 \pm 0.1$.

The $m_{34}$ versus $m_{12}$ distribution in Figure~\ref{fig:resultsDataScatterLowMass} shows that there is no evidence of a signal-like resonance even outside of the $120~\GeV < m_{4\ell} < 130~\GeV$ window applied in this selection: 16 events are observed outside of this mass window, compared to a MC-based prediction of $15 \pm 2$ events from non-resonant SM $ZZ$ processes. These 16 events are shown in the validation region in Figure~\ref{fig:ZdZdLowMassAnalysis_validation}.

\begin{table}[htbp!]
  \begin{center}
    \begin{small}
      \begin{tabular*}{0.4\textwidth}{@{\extracolsep{\fill}}lc}
        \toprule
		\textbf{Process} & Yield \\
        \hline
        $\ZZllll$                                      & $0.10\pm0.01$ \\
        $\HZZllll$                                     & $0.1\pm0.1$   \\
	VVV/VBS                                             & $0.06\pm0.03$ \\
        Heavy flavour                                  & $0.07\pm0.04$ \\
        \hline
        Total                                          & $0.4\pm0.1$   \\
        \hline
        Data                                           & 0             \\
        \bottomrule
      \end{tabular*}
      \caption{\label{tab:ZdZdLowMassAnalysis_yields}Expected event yields of the SM background processes and observed data in the $H\rightarrow XX \rightarrow 4\mu$ ($1~\GeV < m_{X} < 15~\GeV$) selection. The uncertainties include MC-statistical and systematic components (systematic uncertainties are discussed in Section~\ref{sec:ZdZdLowMassAnalysis_Results}).
      }
    \end{small}
  \end{center}
\end{table}

\section{Interpretation and discussion}
\label{sec:Discussion_on_Interpretation}

The results %
do not show evidence for the signal processes of $H\rightarrow ZX \rightarrow 
4\ell$ or $H\rightarrow XX \rightarrow 4\ell$.
The results are therefore interpreted in terms of limits on the benchmark models 
presented in Section~\ref{sec:theory}.

For the $H\rightarrow ZX \rightarrow 4\ell$ analysis, the signal shape is 
obtained directly from simulation using the $Z_{(d)}Z_d$ benchmark  
model~\cite{Curtin:2014cca,Curtin:2013fra}. For the $H\rightarrow XX 
\rightarrow 4\ell$ analysis, a simple Gaussian model is used for a generic 
signal in the $\langle m_{\ell\ell}\rangle$ observable, with the mean and 
standard deviation depending on the mass scale and resolution, respectively, in 
each decay channel. These scales and resolutions are estimated directly from 
simulation. The mass scale is found to have a $-2$\% bias in $X\rightarrow ee$ 
decays, and $-0.5$\% bias in $X\rightarrow\mu\mu$ decays (i.e.\ a $-1$\% bias is 
used in the $ee\mu\mu$ channel). The mass resolutions are estimated to be 3.5\% 
in $X\rightarrow ee$ decays and 1.9\% in $X\rightarrow \mu\mu$ decays (meaning, 
for example, that the standard deviation in the $4\mu$ channel is about 
$1/\sqrt{2} \times 1.9\% \approx 1.34\%$ of $m_{X}$). These scales and 
resolutions are valid across the full mass range considered ($1$--$60$~\GeV).

\subsection{Limits on fiducial cross-sections}
\label{sec:fiducialLimits}
\begin{figure}[!htbp]
	\begin{center}
		\subfloat[Efficiencies\label{fig:efficiency_ZZd}]
		{\includegraphics[width=0.49\textwidth]{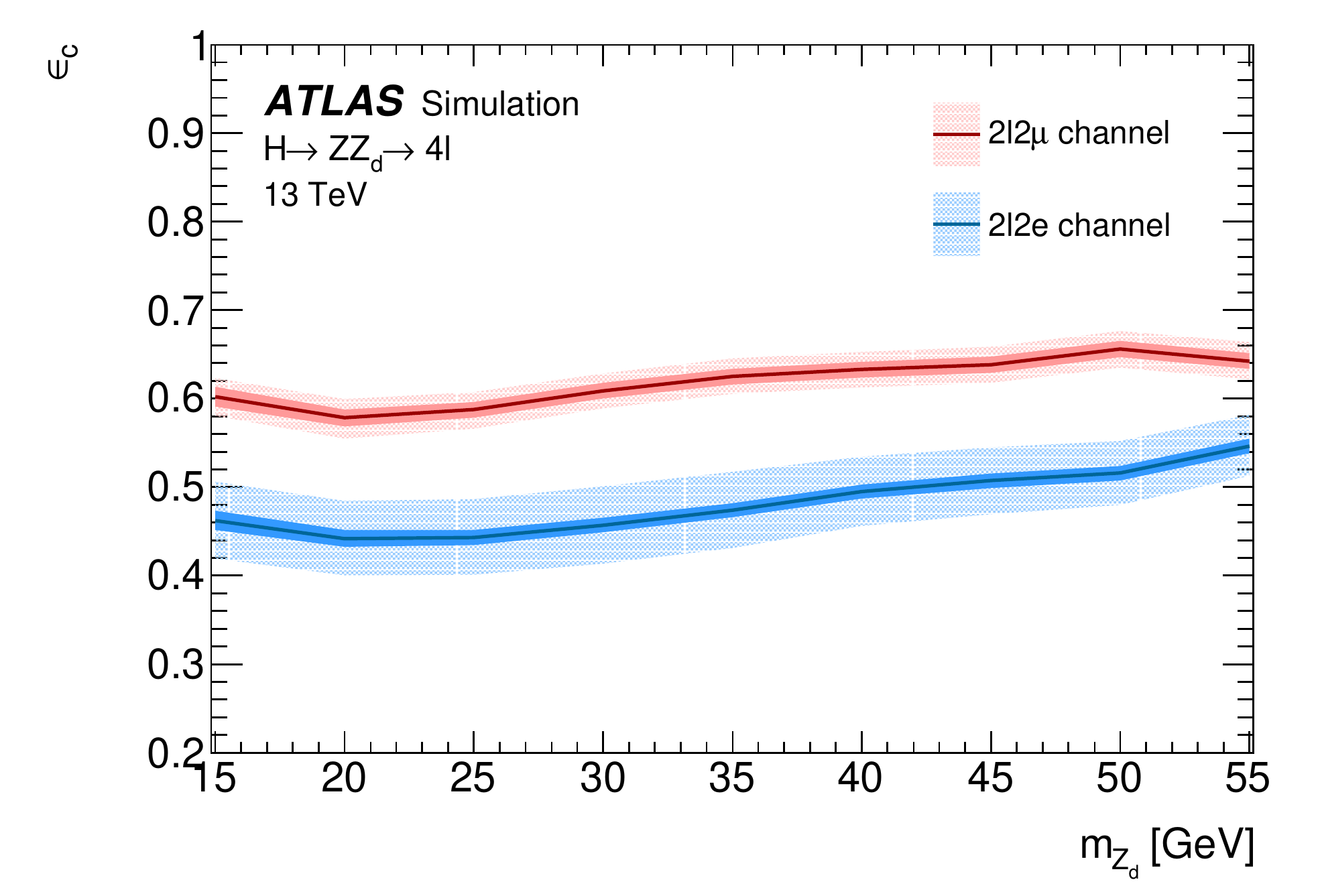}}
		\subfloat[Fiducial cross-sections\label{fig:fiducialLimits_ZZd}]
		{\includegraphics[width=0.49\textwidth]{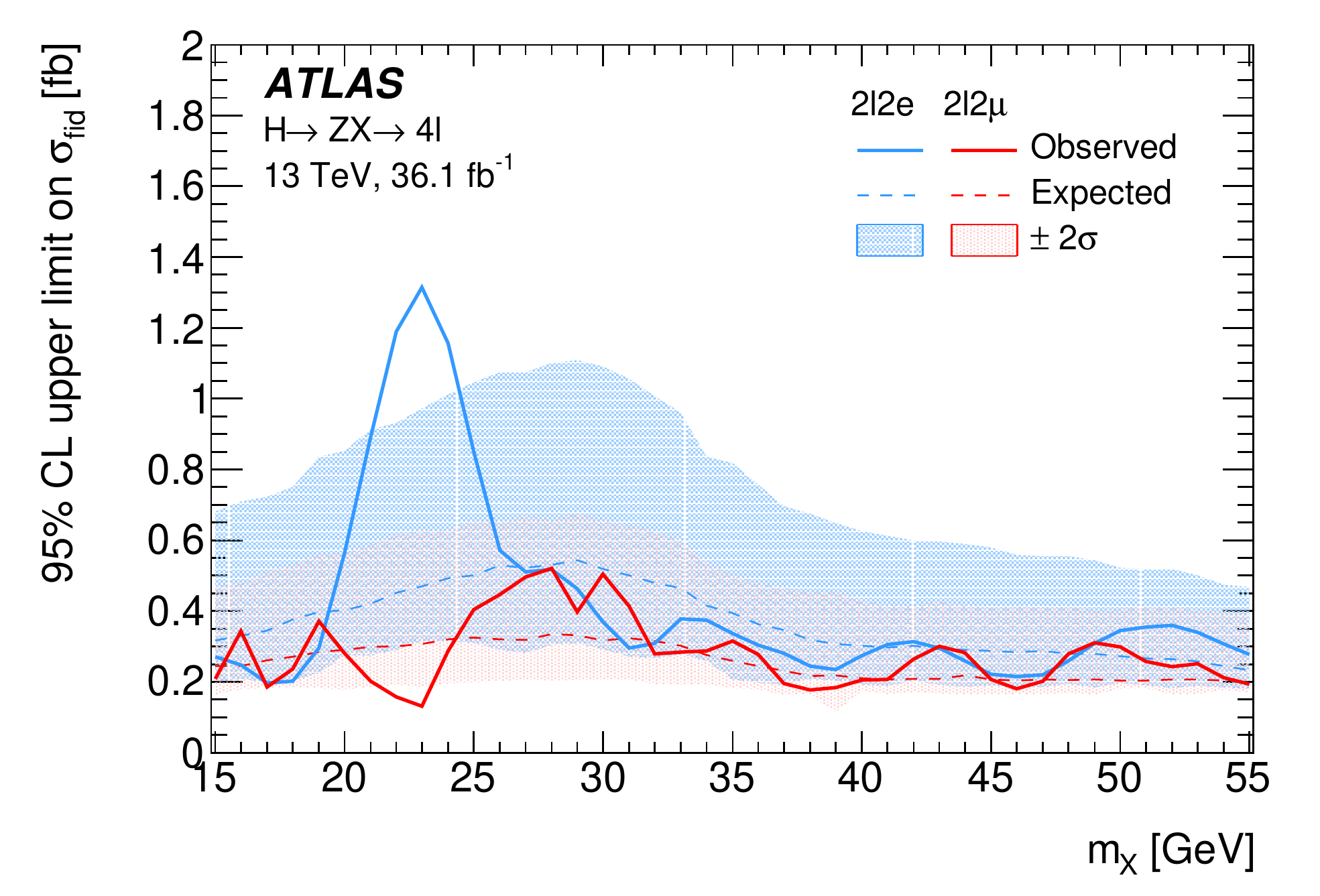}}
		\caption{(a) Per-channel efficiencies $\epsilon_c$ calculated in the 
		fiducial volume described in the $H\rightarrow ZX\rightarrow 4\ell$~ 
		column of Table~\ref{tab:fiducial_selection}. The dark band is the statistical uncertainty and the lighter band is the systematic uncertainty. These efficiencies were computed using the $H\rightarrow 
		ZZ_d\rightarrow 4\ell$ model. (b) Upper limits at the 95\% CL on fiducial cross-sections for the $H\rightarrow ZX \rightarrow 4\ell$ process. The limits from the $H\rightarrow Za\rightarrow 4\ell$ search are valid only for the $2\ell2\mu$ channel as the $H\rightarrow Za$ model assumes $\BR(a\rightarrow \mu\mu)=100\%$.   \label{fig:fiducialLimits}}
	\end{center}
\end{figure}

\begin{figure}[!htbp]
	\begin{center}
		\subfloat[Efficiencies\label{fig:efficiency_ZdZd}]{
			\includegraphics[width=0.49\textwidth]{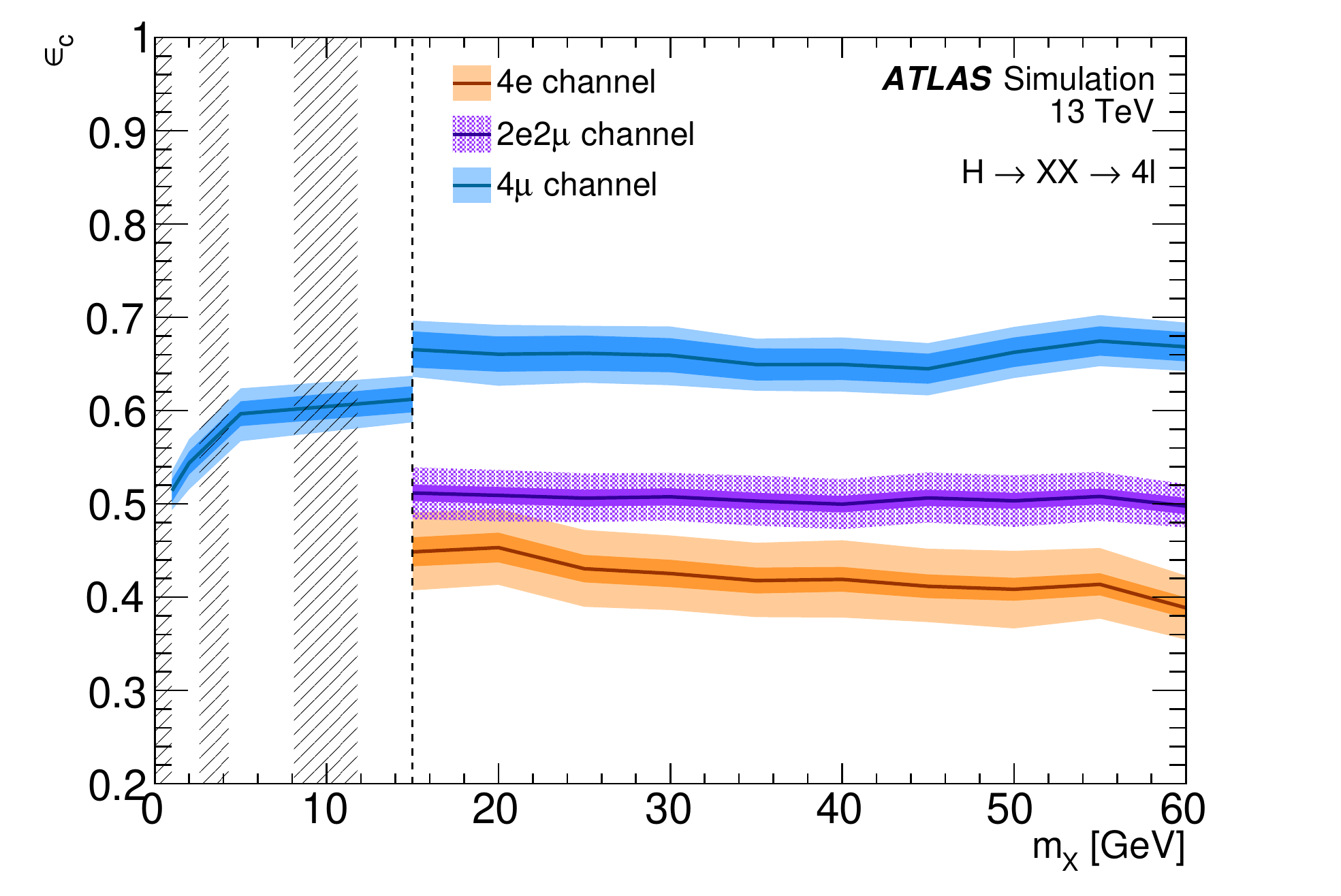}}
		\subfloat[Fiducial cross-sections\label{fig:fiducialLimits_ZdZd}]
		{\includegraphics[width=0.49\textwidth]{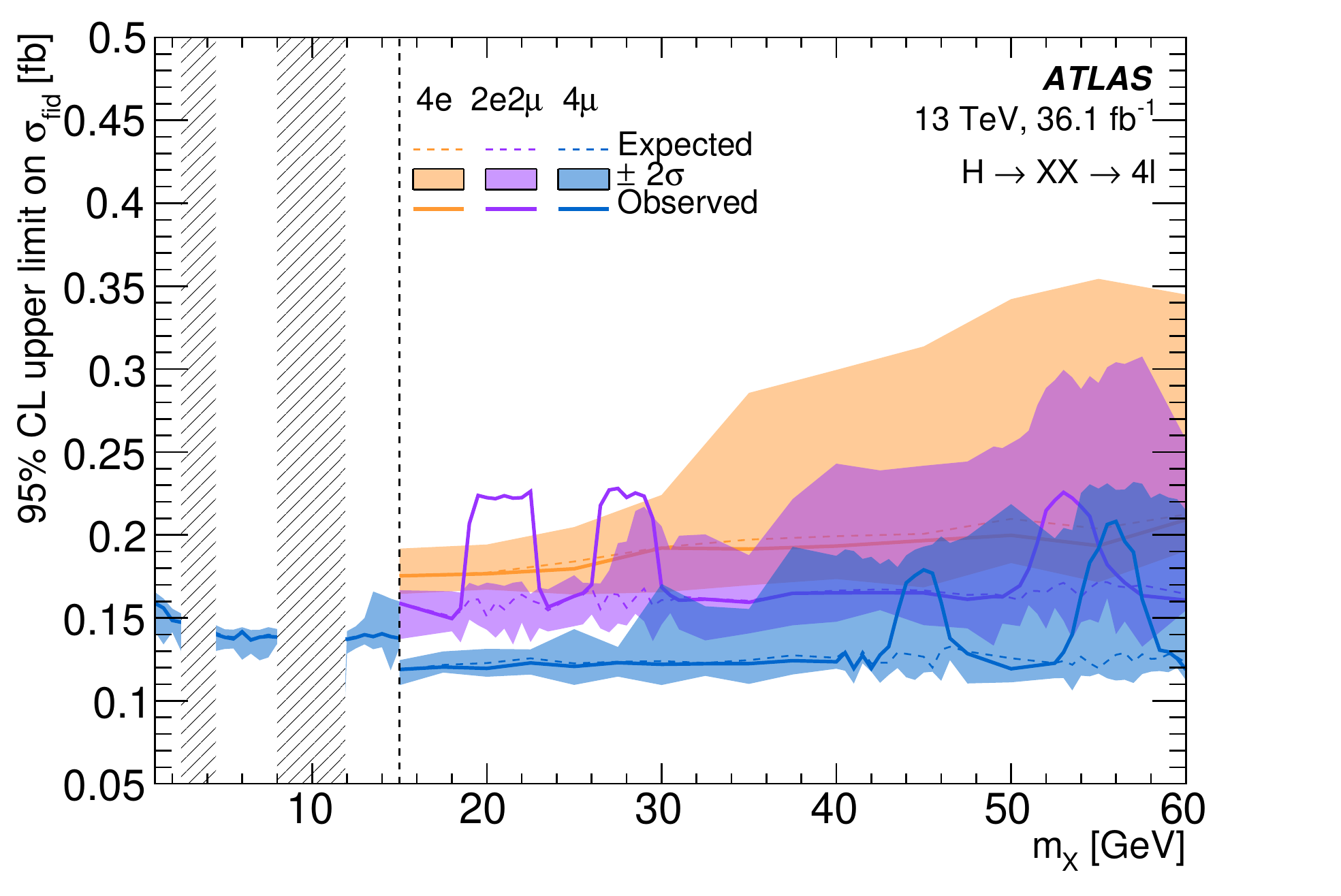}}
		\caption{(a) Model-independent per-channel efficiencies $\epsilon_c$ 
		calculated in the 
		fiducial volumes described in the $1~\GeV < m_{X} < 15~\GeV$ and 
		$15~\GeV < m_{X} < 60~\GeV$ columns of 
		Table~\ref{tab:fiducial_selection} (i.e.\ separate phase spaces are 
		defined for $m_{X}$ above and below $15~\GeV$). The dark band is the statistical uncertainty and the lighter band is the systematic uncertainty. 
		(b) Upper limits at the 95\% CL on fiducial cross-sections for the for the $H\rightarrow XX \rightarrow 4\ell$ process. The step change in the fiducial cross-section limit in the $4\mu$ channel is due to the change in efficiency
			caused by the change in fiducial phase-space definition. The shaded areas are the quarkonia veto regions.  \label{fig:fiducialLimits}}
	\end{center}
\end{figure}

The results were interpreted as limits on fiducial cross-sections by estimating 
the reconstruction efficiencies of each channel $\epsilon_c$ in 
the fiducial phase spaces 
defined in Table~\ref{tab:fiducial_selection}. The fiducial selections were 
chosen to mimic the analysis selections described in 
Sections~\ref{sec:ZZdAnalysis_EventSelection} 
and \ref{sec:ZdZdHighMassAnalysis_EventSelection}. The leptons are ``dressed'', 
i.e.\  in order to emulate the effects of quasi-collinear electromagnetic 
radiation from the charged leptons on their experimental reconstruction in the 
detector~\cite{ATL-PHYS-PUB-2015-013}, the four-momenta of all prompt photons 
within $\Delta R = 0.1$ of a lepton are added to the four-momentum of the 
closest lepton. For the $H\rightarrow XX$ search the efficiencies (shown in 
figure~\ref{fig:efficiency_ZdZd}) are estimated with the $Z_dZ_d$ benchmark 
model and were verified to be compatible with the $aa$ benchmark model to 
within $3\%$ across the whole mass range. For the $ZX$ search the efficiencies 
(shown in figure~\ref{fig:efficiency_ZZd}) are estimated with the $ZZ_d$ 
benchmark model, but no verification was explicitly made to confirm if these 
efficiencies are valid for a $Za$ process.  
Assuming that the verified compatibility between efficiencies of $Z_dZ_d$ and 
$aa$ 
processes applies equally to $ZZ_d$ and $Za$ processes, upper limits on the 
cross-sections corresponding to these fiducial 
phase spaces %
should be applicable to any models of $125~\GeV$ Higgs boson decays to four 
leptons via one (with an associated Z boson) or two intermediate, on-shell, narrow, promptly decaying 
bosons. The fiducial requirements are applied to the four leptons in this decay.  These 
efficiencies are used to compute 95\% CL upper limits on the cross-sections in 
the fiducial phase spaces defined for the $H\rightarrow ZX \rightarrow 4\ell$ 
and $H\rightarrow XX \rightarrow 4\ell$ searches. These model-independent 
limits are computed using  the CL$_{\text{s}}$ frequentist 
formalism~\cite{Read:2002hq} with the profile-likelihood-ratio test 
statistic~\cite{Cowan:2010js} (systematics are represented with nuisance 
parameters which are then profiled in the calculation of the test statistic). 
The results are shown  in Figures \ref{fig:fiducialLimits_ZZd} and 
\ref{fig:fiducialLimits_ZdZd}, respectively. 
Impact of the systematic uncertainties on the limits is small. 

For the $H\rightarrow ZX\rightarrow 4\ell$ search, a local excess of $3\sigma$ at $m_{Z_d}=23$~\GeV~is observed in the $2\ell 2e$ channel.  However, the total observed and predicted event counts in this channel agree within $0.5\sigma$. No local excess is observed in the $2\ell2\mu$ channel. 

The width of the $2\sigma$ expected limit bands of 
Figure~\ref{fig:fiducialLimits_ZdZd} increases towards large values of $m_X$  
because more events are expected from background-only processes at this end of 
the mass spectrum; the larger expected background leads to a greater spread 
in the limits obtained with pseudoexperiments generated with the background-only 
hypothesis.

\begin{table}[!htbp]
        {\footnotesize
        \begin{center}
         \begin{tabular}{l|>{\centering\arraybackslash}p{0.26\linewidth}|>{\centering\arraybackslash}p{0.26\linewidth}|>{\centering\arraybackslash}p{0.26\linewidth}}
                \toprule
                & \textbf{$H\rightarrow ZX \rightarrow 4\ell$ \newline ($15~\GeV < m_{X} < 55~\GeV$)} &
                \textbf{$H\rightarrow XX \rightarrow 4\ell$ \newline ($15~\GeV < m_{X} < 60~\GeV$)} &
                \textbf{$H\rightarrow XX \rightarrow 4\mu$ \newline ($1~\GeV  < m_{X} < 15~\GeV$)} \\
                \hline
                
                Electrons &  \multicolumn{3}{c}{Dressed with prompt photons within $\Delta R = 0.1$}\\
                &  \multicolumn{3}{c}{$\pt > 7~\GeV$}\\
                &  \multicolumn{3}{c}{$|\eta|<2.5$}\\
                \hline
                Muons & \multicolumn{3}{c}{Dressed with prompt photons within $\Delta R = 0.1$}\\
                &  \multicolumn{3}{c}{ $\pt > 5~\GeV$ }\\
                &  \multicolumn{3}{c}{$|\eta|<2.7$}\\
                \hline
                Quadruplet & \multicolumn{3}{c}{Three leading-\pt{} leptons satisfy $\pt > 20$~\GeV{}, 15~\GeV{}, 10~\GeV{}}\\
                \cline{2-4}
                & \multicolumn{2}{c|}{$\Delta R > 0.1$ (0.2) between SF (OF) leptons} & - \\
                \cline{2-4}
                & $50~\GeV < m_{12} < 106$~\GeV  & \multicolumn{2}{c}{ $m_{34}/m_{12} > 0.85$ }  \\
                \cline{3-4}
                & $12~\GeV < m_{34} < 115$~\GeV  & $10~\GeV < m_{12,34} < 64$~\GeV{} &  $0.88~\GeV < m_{12,34} < 20~\GeV$ \\
                & $115~\GeV <m_{4\ell} < 130$~\GeV \newline $m_{12,34,14,32}>5 
                $~\GeV & $5~\GeV < m_{14,32} < 75~\GeV$ if $4e$ or $4\mu$ & \\ 
                \cline{3-4}
                &  & 
                \multicolumn{2}{>{\centering\arraybackslash}p{0.52\linewidth}}{ 
                Reject event if either of: $(m_{J/\psi}-0.25~\GeV) < 
                m_{12,34,14,32} < (m_{\psi(2S)}+0.30~\GeV)$ 
                $(m_{\Upsilon(1S)}-0.70~\GeV) < m_{12,34,14,32} < 
                (m_{\Upsilon(3S)}+0.75~\GeV)$}  \\

                \bottomrule
        \end{tabular}

        \end{center}
}
        \caption{Summary of the fiducial phase-space definitions used in this analysis, appropriate for processes of the form $H \rightarrow ZZ_{d} \rightarrow 4\ell$ aand $H \rightarrow XX \rightarrow 4\ell$, where $X$ is a  promptly decaying, on-shell, narrow resonance.}

        \label{tab:fiducial_selection}
\end{table}

\subsection{Limits on branching ratios}

\begin{figure}[!htbp]
	\begin{center}
		\includegraphics[width=0.49\textwidth]{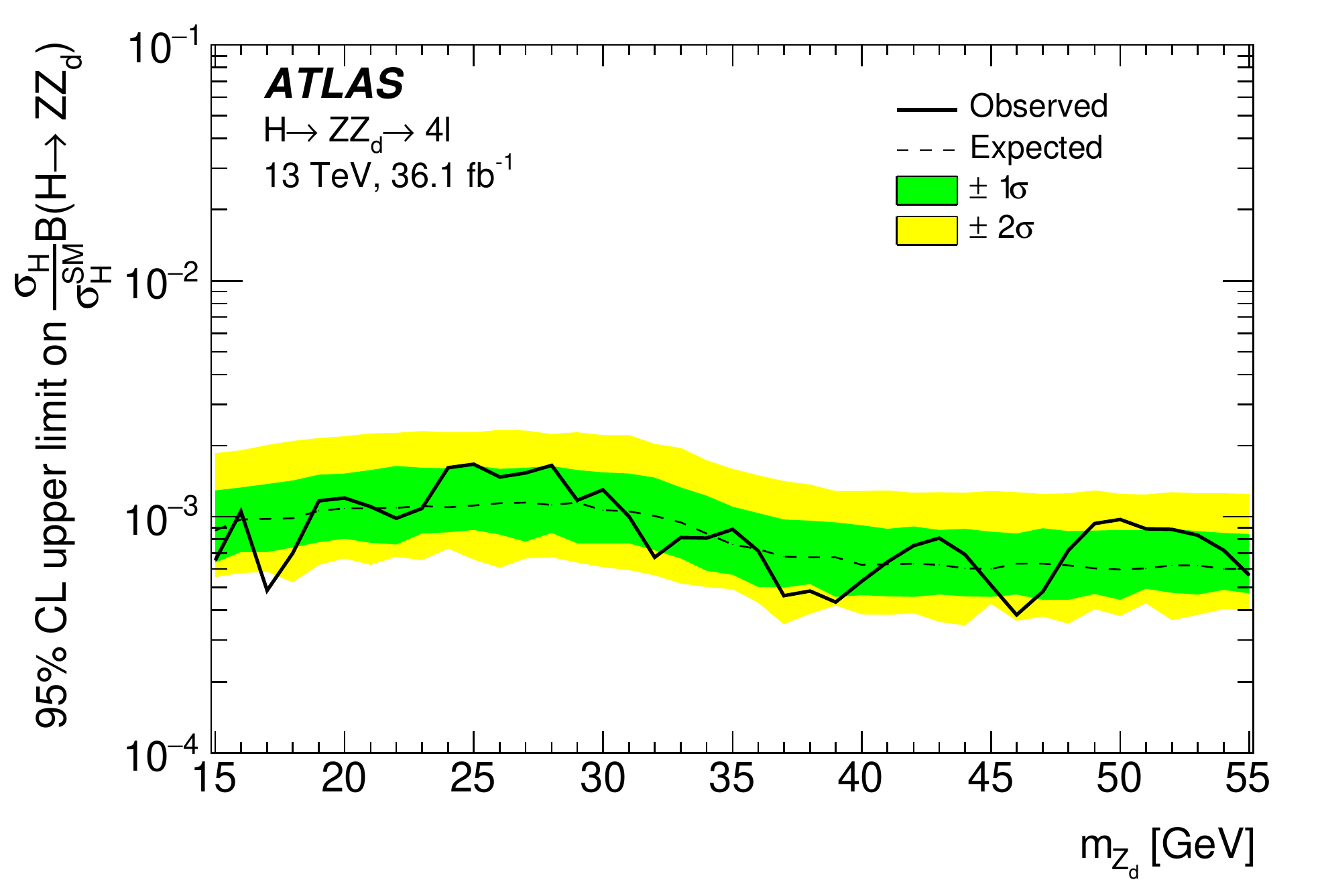}
		\caption{Upper limit at 95\% CL on the branching ratio  for the $H\rightarrow ZZ_d$ process. \label{fig:brLimits_ZZd}}
	\end{center}
\end{figure}

\begin{figure}[!htbp]
	\begin{center}
		\subfloat[$H\rightarrow \Zp\Zp$\label{fig:brLimits2}]{
			\includegraphics[width=0.49\textwidth]{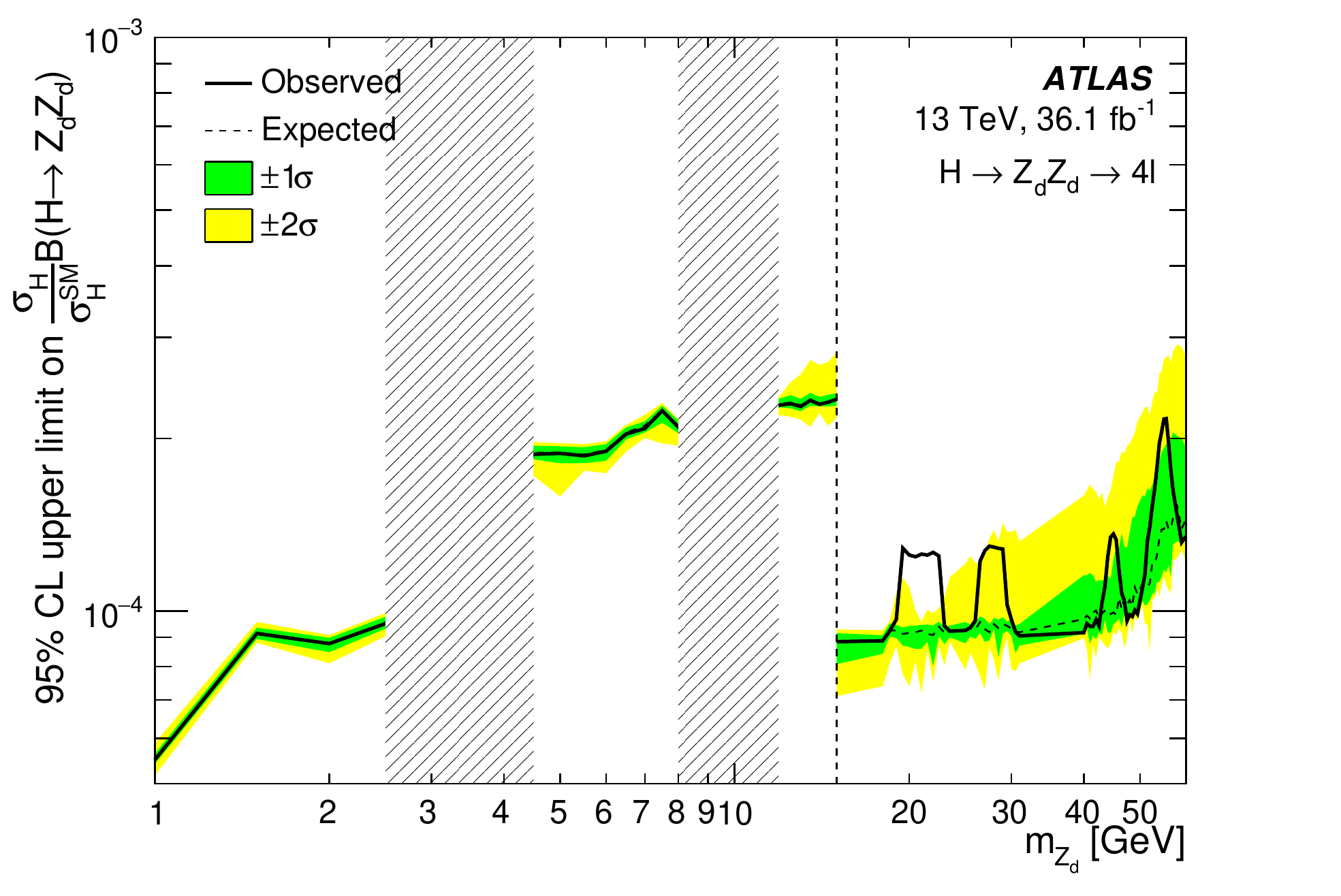}}
		\subfloat[$H\rightarrow aa$\label{fig:brLimits1}]{
			\includegraphics[width=0.49\textwidth]{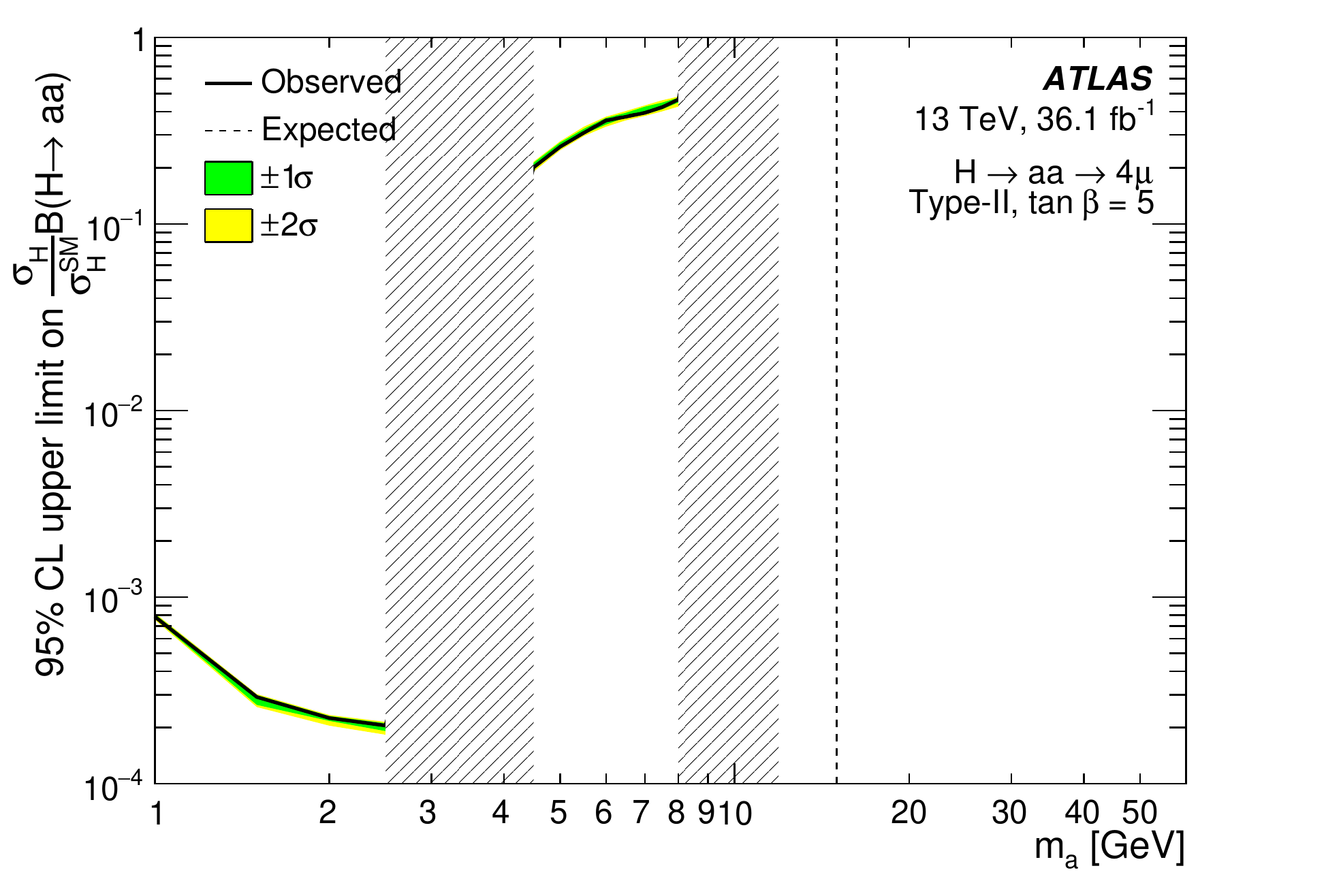}}
    \caption{Upper limit at 95\% CL on the branching ratios for processes (a) $H\rightarrow\Zp\Zp$ and (b) $H\rightarrow aa$, for the two benchmark models studied in this paper. The limit on $H\rightarrow aa$ is greater than 1 for $m_{a}>15~\GeV$. The step change in (a) at the $m_{X}=15~\GeV$ boundary is due to the addition of sensitivity to 4e and $2e2\mu$ final states (lowering the limit). The shaded areas are the quarkonia veto regions. \label{fig:brLimits}}
	\end{center}
\end{figure}

Model-dependent acceptances for the fiducial phase spaces are computed per 
channel for the $H\rightarrow ZZ_d \rightarrow 4\ell$ and $H\rightarrow XX 
\rightarrow 4\ell$ searches. The acceptance for the benchmark vector-boson 
model is estimated for both searches, whereas the acceptance for the benchmark 
pseudoscalar-boson model (type-II 2HDM+S model with $\tan{\beta} = 5$) is estimated only for the $H\rightarrow XX \rightarrow 
4\ell$ search. The acceptances are used in a combined statistical model to 
compute upper limits on $\sigma_{H}\times \BR(H\rightarrow 
ZZ_d\rightarrow 4\ell)$
 and $\sigma_{H}\times \BR(H\rightarrow XX\rightarrow 4\ell)$ for each 
 model. The $\Zp$ model assumes partial fractions of 0.25:0.25:0.25:0.25 for 
 the $4e$:$2e2\mu$:$4\mu$:$2\mu2e$ channels, whereas the $a$ model assumes 
 100\% decay to $4\mu$. These cross-section limits are converted into limits on 
 the branching ratios of $H\rightarrow Z\Zp$, $H\rightarrow\Zp\Zp$ and $\Haa$ 
 by using the theoretical branching ratios for
 $\Zp\rightarrow\ell\ell$ and $a\rightarrow\mu\mu$ from each benchmark 
 model~\cite{Curtin:2014cca,Curtin:2013fra}, and assuming for $\sigma_{H}$  the 
 SM cross-section\footnote{This assumes that the presence of BSM decays of the 
 Higgs boson does not signicantly alter the Higgs boson production cross-section 
 from the SM prediction.} for Higgs boson production at 
 $\sqrt{s}=13~\TeV$~\cite{LHCHiggsCrossSectionWorkingGroup:2012vm}. The limits 
 on these branching ratios are shown in Figures~\ref{fig:brLimits_ZZd} and 
 \ref{fig:brLimits} for the $H\rightarrow ZZ_d \rightarrow 4\ell$ and 
 $H\rightarrow XX \rightarrow 4\ell$ searches, respectively. The observed limit 
 for $\BR(H\rightarrow aa)$ (Figure~\ref{fig:brLimits1}) for 
 $m_a>15~\GeV$ is 
 greater than 1 (i.e.\ this search has no sensitivity to this model in that mass 
 range). The limit on the branching ratio for $H\rightarrow Z_dZ_d \rightarrow 4\ell$ improves on the Run~1 
 result of Ref.~\cite{EXOT-2013-15} by about a factor of four, which corresponds 
 to the increase in both luminosity and Higgs boson production cross-section between Run~1 and 
 Run~2.

\FloatBarrier

\section{Conclusion}
\label{sec:conclusion}
Searches are performed for exotic decays of the Standard Model Higgs boson with 
a mass of 125~\GeV~to one or two new spin-1 particles, $H\rightarrow ZZ_d$ 
and $\HZpZp$, or spin-0 particles, $H\rightarrow Za$ and $H\rightarrow aa$, using proton--proton 
collision data produced at $\sqrt{s} = 13~\TeV$ and recorded by the ATLAS 
detector at the LHC in 2015 and 2016.  The data
correspond to a combined integrated luminosity of $36.1~\text{fb}^{-1}$. The 
searches explore the final state consisting of four leptons (electrons or 
muons) produced from the prompt decays of the intermediate boson states. The 
search targeting the $ZX$ intermediate state is sensitive to $Z_d$ masses 
between 15 and 55~\GeV, and the search targeting the $XX$ intermediate 
state is sensitive to $Z_d$ masses between 15 and 60~\GeV. Intermediate states 
of $XX$ ($X=a$ or $Z_d$) where the boson mass $m_{X}$ is between 1 and 15~\GeV 
~(excluding masses near known dilepton resonances from quarkonium states) are searched 
for in events with four muons.

The data are found to be globally consistent with SM background predictions. 
Upper limits (dependent on the mass of the intermediate exotic boson) are set on the 
branching ratio of Higgs boson to $ZZ_d$, $Z_d Z_d$, and 
$aa$, corresponding to $\BR(H\rightarrow ZZ_d)\approx 0.1\%$, 
$\BR(H\rightarrow Z_dZ_d)\approx 0.01\%$, and $\BR(H\rightarrow 
aa)\approx 1\%$ respectively (depending on mass range). These limits are 
obtained under the assumption of 
a SM production cross-section for Higgs bosons, and assume prompt decay of the 
exotic bosons. Limits on fiducial 
cross-sections are also computed, which can be used for testing models 
different from the benchmark models used in this paper.

\section*{Acknowledgements}

We thank CERN for the very successful operation of the LHC, as well as the
support staff from our institutions without whom ATLAS could not be
operated efficiently.

We acknowledge the support of ANPCyT, Argentina; YerPhI, Armenia; ARC, Australia; BMWFW and FWF, Austria; ANAS, Azerbaijan; SSTC, Belarus; CNPq and FAPESP, Brazil; NSERC, NRC and CFI, Canada; CERN; CONICYT, Chile; CAS, MOST and NSFC, China; COLCIENCIAS, Colombia; MSMT CR, MPO CR and VSC CR, Czech Republic; DNRF and DNSRC, Denmark; IN2P3-CNRS, CEA-DRF/IRFU, France; SRNSFG, Georgia; BMBF, HGF, and MPG, Germany; GSRT, Greece; RGC, Hong Kong SAR, China; ISF, I-CORE and Benoziyo Center, Israel; INFN, Italy; MEXT and JSPS, Japan; CNRST, Morocco; NWO, Netherlands; RCN, Norway; MNiSW and NCN, Poland; FCT, Portugal; MNE/IFA, Romania; MES of Russia and NRC KI, Russian Federation; JINR; MESTD, Serbia; MSSR, Slovakia; ARRS and MIZ\v{S}, Slovenia; DST/NRF, South Africa; MINECO, Spain; SRC and Wallenberg Foundation, Sweden; SERI, SNSF and Cantons of Bern and Geneva, Switzerland; MOST, Taiwan; TAEK, Turkey; STFC, United Kingdom; DOE and NSF, United States of America. In addition, individual groups and members have received support from BCKDF, the Canada Council, CANARIE, CRC, Compute Canada, FQRNT, and the Ontario Innovation Trust, Canada; EPLANET, ERC, ERDF, FP7, Horizon 2020 and Marie Sk{\l}odowska-Curie Actions, European Union; Investissements d'Avenir Labex and Idex, ANR, R{\'e}gion Auvergne and Fondation Partager le Savoir, France; DFG and AvH Foundation, Germany; Herakleitos, Thales and Aristeia programmes co-financed by EU-ESF and the Greek NSRF; BSF, GIF and Minerva, Israel; BRF, Norway; CERCA Programme Generalitat de Catalunya, Generalitat Valenciana, Spain; the Royal Society and Leverhulme Trust, United Kingdom.

The crucial computing support from all WLCG partners is acknowledged gratefully, in particular from CERN, the ATLAS Tier-1 facilities at TRIUMF (Canada), NDGF (Denmark, Norway, Sweden), CC-IN2P3 (France), KIT/GridKA (Germany), INFN-CNAF (Italy), NL-T1 (Netherlands), PIC (Spain), ASGC (Taiwan), RAL (UK) and BNL (USA), the Tier-2 facilities worldwide and large non-WLCG resource providers. Major contributors of computing resources are listed in Ref.~\cite{ATL-GEN-PUB-2016-002}.

\printbibliography

\clearpage
 
\begin{flushleft}
{\Large The ATLAS Collaboration}

\bigskip

M.~Aaboud$^\textrm{\scriptsize 34d}$,    
G.~Aad$^\textrm{\scriptsize 99}$,    
B.~Abbott$^\textrm{\scriptsize 124}$,    
O.~Abdinov$^\textrm{\scriptsize 13,*}$,    
B.~Abeloos$^\textrm{\scriptsize 128}$,    
S.H.~Abidi$^\textrm{\scriptsize 165}$,    
O.S.~AbouZeid$^\textrm{\scriptsize 143}$,    
N.L.~Abraham$^\textrm{\scriptsize 153}$,    
H.~Abramowicz$^\textrm{\scriptsize 159}$,    
H.~Abreu$^\textrm{\scriptsize 158}$
% ...
\bigskip
\\

% ...
$^{13}$Institute of Physics, Azerbaijan Academy of Sciences, Baku; Azerbaijan.\\
% ...
$^{34}$$^{(a)}$Facult\'e des Sciences Ain Chock, R\'eseau Universitaire de Physique des Hautes Energies - Universit\'e Hassan II, Casablanca;$^{(b)}$Centre National de l'Energie des Sciences Techniques Nucleaires (CNESTEN), Rabat;$^{(c)}$Facult\'e des Sciences Semlalia, Universit\'e Cadi Ayyad, LPHEA-Marrakech;$^{(d)}$Facult\'e des Sciences, Universit\'e Mohamed Premier and LPTPM, Oujda;$^{(e)}$Facult\'e des sciences, Universit\'e Mohammed V, Rabat; Morocco.\\
% ...
$^{99}$CPPM, Aix-Marseille Universit\'e, CNRS/IN2P3, Marseille; France.\\
% ...
$^{124}$Homer L. Dodge Department of Physics and Astronomy, University of Oklahoma, Norman OK; United States of America.\\
% ...
$^{128}$LAL, Universit\'e Paris-Sud, CNRS/IN2P3, Universit\'e Paris-Saclay, Orsay; France.\\
% ...
$^{143}$Santa Cruz Institute for Particle Physics, University of California Santa Cruz, Santa Cruz CA; United States of America.\\
% ...
$^{153}$Department of Physics and Astronomy, University of Sussex, Brighton; United Kingdom.\\
% ...
$^{158}$Department of Physics, Technion, Israel Institute of Technology, Haifa; Israel.\\
$^{159}$Raymond and Beverly Sackler School of Physics and Astronomy, Tel Aviv University, Tel Aviv; Israel.\\
% ...
$^{165}$Department of Physics, University of Toronto, Toronto ON; Canada.\\
% ...
$^{a}$ Also at  Department of Physics, University of Malaya, Kuala Lumpur; Malaysia.\\
$^{b}$ Also at Borough of Manhattan Community College, City University of New York, NY; United States of America.\\
$^{c}$ Also at Centre for High Performance Computing, CSIR Campus, Rosebank, Cape Town; South Africa.\\
$^{d}$ Also at CERN, Geneva; Switzerland.\\
$^{e}$ Also at CPPM, Aix-Marseille Universit\'e, CNRS/IN2P3, Marseille; France.\\
% ...
$^{*}$ Deceased

\end{flushleft}


\end{document}